\input harvmac
\input epsf

%
%
\def\nextline{\hfil\break}

\noblackbox
%


\def\unlockat{\catcode`\@=11}
\def\lockat{\catcode`\@=12}

\unlockat

\def\newsec#1{\global\advance\secno by1\message{(\the\secno. #1)}
\global\subsecno=0\global\subsubsecno=0\eqnres@t\noindent
{\bf\the\secno. #1}
\writetoca{{\secsym} {#1}}\par\nobreak\medskip\nobreak}
\global\newcount\subsecno \global\subsecno=0
\def\subsec#1{\global\advance\subsecno
by1\message{(\secsym\the\subsecno. #1)}
\ifnum\lastpenalty>9000\else\bigbreak\fi\global\subsubsecno=0
\noindent{\it\secsym\the\subsecno. #1}
\writetoca{\string\quad {\secsym\the\subsecno.} {#1}}
\par\nobreak\medskip\nobreak}
\global\newcount\subsubsecno \global\subsubsecno=0
\def\subsubsec#1{\global\advance\subsubsecno by1
\message{(\secsym\the\subsecno.\the\subsubsecno. #1)}
\ifnum\lastpenalty>9000\else\bigbreak\fi
\noindent\quad{\secsym\the\subsecno.\the\subsubsecno.}{#1}
\writetoca{\string\qquad{\secsym\the\subsecno.\the\subsubsecno.}{#1}}
\par\nobreak\medskip\nobreak}

\def\subsubseclab#1{\DefWarn#1\xdef
#1{\noexpand\hyperref{}{subsubsection}%
{\secsym\the\subsecno.\the\subsubsecno}%
{\secsym\the\subsecno.\the\subsubsecno}}%
\writedef{#1\leftbracket#1}\wrlabeL{#1=#1}}
\lockat

\def\hone{h^{(1)} }


\def\CN {{\cal N}}

\def\CO {{\cal O}}

\def\CE{{\cal E }}
\def\CV{{\cal V }}

\def\CS {{\cal S }}

\font\manual=manfnt \def\dbend{\lower3.5pt\hbox{\manual\char127}}

\def\IZ{\relax\ifmmode\mathchoice
{\hbox{\cmss Z\kern-.4em Z}}{\hbox{\cmss Z\kern-.4em Z}}
{\lower.9pt\hbox{\cmsss Z\kern-.4em Z}}
{\lower1.2pt\hbox{\cmsss Z\kern-.4em Z}}\else{\cmss Z\kern-.4em
Z}\fi}
\def\half{{1\over 2}}

\def\p{\partial}

\def\CN {{\cal N}}

\def\CO {{\cal O}}

\def\CE{{\cal E }}
\def\CV{{\cal V }}

\def\CS {{\cal S }}


\def\IZ{\relax\ifmmode\mathchoice
{\hbox{\cmss Z\kern-.4em Z}}{\hbox{\cmss Z\kern-.4em Z}}
{\lower.9pt\hbox{\cmsss Z\kern-.4em Z}}
{\lower1.2pt\hbox{\cmsss Z\kern-.4em Z}}\else{\cmss Z\kern-.4em
Z}\fi}
\def\IB{\relax{\rm I\kern-.18em B}}
\def\IC{{\relax\hbox{$\inbar\kern-.3em{\rm C}$}}}
\def\ID{\relax{\rm I\kern-.18em D}}
\def\IE{\relax{\rm I\kern-.18em E}}
\def\IF{\relax{\rm I\kern-.18em F}}
\def\IG{\relax\hbox{$\inbar\kern-.3em{\rm G}$}}
\def\IGa{\relax\hbox{${\rm I}\kern-.18em\Gamma$}}
\def\IH{\relax{\rm I\kern-.18em H}}
\def\II{\relax{\rm I\kern-.18em I}}
\def\IK{\relax{\rm I\kern-.18em K}}
\def\IP{\relax{\rm I\kern-.18em P}}
\def\IQ{\relax\hbox{$\inbar\kern-.3em{\rm Q}$}}

\def\inbar{\,\vrule height1.5ex width.4pt depth0pt}

\def\mod{{\rm mod}}
\def\p{\partial}

\font\cmss=cmss10 \font\cmsss=cmss10 at 7pt
\def\IR{\relax{\rm I\kern-.18em R}}

%
%



\def\sector#1#2{\ {\scriptstyle #1}\hskip 1mm
\mathop{\opensquare}\limits_{\lower 1mm\hbox{$\scriptstyle#2$}}\hskip 1mm}

\def\tsector#1#2{\ {\scriptstyle #1}\hskip 1mm
\mathop{\opensquare}\limits_{\lower 1mm\hbox{$\scriptstyle#2$}}^\sim\hskip 1mm}


\def\inbar{\,\vrule height1.5ex width.4pt depth0pt}

\def\p{\partial}

\font\cmss=cmss10 \font\cmsss=cmss10 at 7pt
\def\IR{\relax{\rm I\kern-.18em R}}


\def\frac#1#2{{#1\over#2}}

\def\half{\frac12}

\def\inbar{\,\vrule height1.5ex width.4pt depth0pt}
\def\IC{\relax\hbox{$\inbar\kern-.3em{\rm C}$}}
\def\IR{\relax{\rm I\kern-.18em R}}
\def\IP{\relax{\rm I\kern-.18em P}}

%
%
\catcode`\@=11
\def\slash#1{\mathord{\mathpalette\c@ncel{#1}}}
\overfullrule=0pt

\def\CC{{\cal C}}

\def\II{{\cal I}}

\def\NN{{\cal N}}
\def\OO{{\cal O}}

\def\underrel#1\over#2{\mathrel{\mathop{\kern\z@#1}\limits_{#2}}}

\catcode`\@=12


%

\def\det{{\rm det}}
\def\tr{{\rm tr}}
\def\mod{{\rm mod}}

\def\det{{\rm det}}
\def\exp{{\rm exp}}


\def\ra{{\rightarrow}}
\def\HJ{{\rm Hirzebruch-Jung}}
\def\mn{{\rm mod}\; n}
\def\Xbar{{\bar X}}
\def\eff{{\rm eff}}


 \def\p{\partial}
\def\a{\alpha}
\def\b{\beta}
\def\e{\epsilon}
\def\CS{{\cal S}}
\def\CV{{\cal V}}
\def\mod{{\rm mod}}

\def\re{{\rm Re}}

 \def\p{\partial}
\def\a{\alpha}
\def\b{\beta}
\def\e{\epsilon}
\def\CE{{\cal E}}
\def\CS{{\cal S}}
\def\CU{{\cal U}}
\def\CV{{\cal V}}
\def\mod{{\rm mod}}

\def\re{{\rm Re}}
\def\I{{\cal I}}
\def\It{{\tilde {\cal I}}}

%



\lref\aps{
A.~Adams, J.~Polchinski and E.~Silverstein,
``Don't panic! Closed string tachyons in ALE space-times,''
JHEP {\bf 0110}, 029 (2001)
[arXiv:hep-th/0108075].
}

\lref\hkmm{
J.~A.~Harvey, D.~Kutasov, E.~J.~Martinec and G.~Moore,
``Localized tachyons and RG flows,''
arXiv:hep-th/0111154.
}

\lref\DixonQV{
L.~J.~Dixon, D.~Friedan, E.~J.~Martinec and S.~H.~Shenker,
``The Conformal Field Theory Of Orbifolds,''
Nucl.\ Phys.\ B {\bf 282}, 13 (1987).
}

\lref\mm{
E.~J.~Martinec and G.~Moore,
``On decay of K-theory,''
arXiv:hep-th/0212059.
}

\lref\VafaRA{
C.~Vafa,
``Mirror symmetry and closed string tachyon condensation,''
arXiv:hep-th/0111051.
}

\lref\mt{
S.~Minwalla and T.~Takayanagi,
``Evolution of D-branes under closed string tachyon condensation,''
JHEP {\bf 0309}, 011 (2003)
[arXiv:hep-th/0307248].
}

\lref\DavidVM{
J.~R.~David, M.~Gutperle, M.~Headrick and S.~Minwalla,
``Closed string tachyon condensation on twisted circles,''
JHEP {\bf 0202}, 041 (2002)
[arXiv:hep-th/0111212].
}

\lref\DouglasHQ{
M.~R.~Douglas and B.~Fiol,
``D-branes and discrete torsion. II,''
arXiv:hep-th/9903031.
}
\lref\BerkoozIS{
M.~Berkooz and M.~R.~Douglas,
``Five-branes in M(atrix) theory,''
Phys.\ Lett.\ B {\bf 395}, 196 (1997)
[arXiv:hep-th/9610236].
}
\lref\BerkoozKM{
M.~Berkooz, M.~R.~Douglas and R.~G.~Leigh,
``Branes intersecting at angles,''
Nucl.\ Phys.\ B {\bf 480}, 265 (1996)
[arXiv:hep-th/9606139].
}
\lref\bcr{
M.~Billo, B.~Craps and F.~Roose,
``Orbifold boundary states from Cardy's condition,''
JHEP {\bf 0101}, 038 (2001)
[arXiv:hep-th/0011060].
}

\lref\dm{
M.~R.~Douglas and G.~W.~Moore,
``D-branes, Quivers, and ALE Instantons,''
arXiv:hep-th/9603167.
}

\lref\HoriCK{
K.~Hori, A.~Iqbal and C.~Vafa,
``D-branes and mirror symmetry,''
arXiv:hep-th/0005247.
}

\lref\HoriIC{
K.~Hori,
``Linear models of supersymmetric D-branes,''
arXiv:hep-th/0012179.
}

\lref\MaldacenaSN{
  J.~Maldacena, G.~W.~Moore, N.~Seiberg and D.~Shih,
  ``Exact vs. semiclassical target space of the minimal string,''
  JHEP {\bf 0410}, 020 (2004)
  [arXiv:hep-th/0408039].
}

\lref\MartinecWG{
E.~J.~Martinec and G.~Moore,
``On decay of K-theory,''
arXiv:hep-th/0212059.
}

\lref\MartinecTZ{
E.~J.~Martinec,
``Defects, decay, and dissipated states,''
arXiv:hep-th/0210231.
}


\lref\fulton{W. Fulton, {\it Introduction to Toric Varieties},
Annals of Mathematics Studies, vol. 131; Princeton Univ. Press (1993).} 
\lref\bpv{W.~Barth, C.~Peters, A.~Van de Ven, {\it Compact Complex Surfaces};
Springer-Verlag (1984).} 
\lref\stevens{J. Stevens, ``On the versal deformation
of cyclic quotient singularities'',
in {\it Singularity theory and its applications, part I},
LNM 1462 pp.302-319.}
\lref\ishii{A. Ishii, ``On McKay correspondence
for a finite small subgroup of GL(2,C)'',
to appear in J. Reine Ang. Math.
(available at \nextline
http://www.kusm.kyoto-u.ac.jp/preprint/preprint2000.html).}
\lref\wunram{J. Wunram, ``Reflexive modules on quotient surface
singularities'', Math. Ann. {\bf 279}, 583 (1988).}
\lref\riemenschneider{O. Riemenschneider,
``Special representations and the two-dimensional McKay correspondence''
(available at \nextline
http://www.math.uni-hamburg.de/home/riemenschneider/hokmckay.ps).}
%
\lref\AdamsSV{
A.~Adams, J.~Polchinski and E.~Silverstein,
``Don't panic! Closed string tachyons in ALE space-times,''
JHEP {\bf 0110}, 029 (2001)
arXiv:hep-th/0108075.
}
%
\lref\HarveyWM{
J.~A.~Harvey, D.~Kutasov, E.~J.~Martinec and G.~Moore,
``Localized tachyons and RG flows,''
arXiv:hep-th/0111154.
}
%
\lref\VafaRA{
C.~Vafa,
``Mirror symmetry and closed string tachyon condensation,''
arXiv:hep-th/0111051.
}
%
\lref\MartinecTZ{
E.~J.~Martinec,
``Defects, decay, and dissipated states,''
arXiv:hep-th/0210231.
}
%
\lref\MorrisonFR{
D.~R.~Morrison and M.~Ronen Plesser,
``Summing the instantons: Quantum cohomology 
and mirror symmetry in toric varieties,''
Nucl.\ Phys.\ B {\bf 440}, 279 (1995)
hep-th/9412236.
}
%

\lref\CecottiVB{
S.~Cecotti and C.~Vafa,
``Exact results for supersymmetric sigma models,''
Phys.\ Rev.\ Lett.\  {\bf 68}, 903 (1992)
hep-th/9111016.
}
%
\lref\WittenYC{
E.~Witten,
``Phases of N = 2 theories in two dimensions,''
Nucl.\ Phys.\ B {\bf 403}, 159 (1993)
hep-th/9301042.
}
%
\lref\HoriCK{
K.~Hori, A.~Iqbal and C.~Vafa,
``D-branes and mirror symmetry,''
arXiv:hep-th/0005247.
}
%
\lref\HoriKT{
K.~Hori and C.~Vafa,
``Mirror symmetry,''
arXiv:hep-th/0002222.
}
%
\lref\HoriFJ{
K.~Hori,
``Mirror symmetry and some applications,''
arXiv:hep-th/0106043.
}

\lref\HoriIC{
K.~Hori,
``Linear models of supersymmetric D-branes,''
arXiv:hep-th/0012179.
}
%
\lref\HellermanBU{
S.~Hellerman, S.~Kachru, A.~E.~Lawrence and J.~McGreevy,
``Linear sigma models for open strings,''
JHEP {\bf 0207}, 002 (2002)
arXiv:hep-th/0109069.
}
%
\lref\LercheUY{
W.~Lerche, C.~Vafa and N.~P.~Warner,
``Chiral Rings In N=2 Superconformal Theories,''
Nucl.\ Phys.\ B {\bf 324}, 427 (1989).
}
%
\lref\DouglasSW{
M.~R.~Douglas and G.~W.~Moore,
``D-branes, Quivers, and ALE Instantons,''
arXiv:hep-th/9603167.
}
%
\lref\HarveyNA{
J.~A.~Harvey, D.~Kutasov and E.~J.~Martinec,
``On the relevance of tachyons,''
arXiv:hep-th/0003101.
}
%
\lref\SenMD{
A.~Sen,
``Supersymmetric world-volume action for non-BPS D-branes,''
JHEP {\bf 9910}, 008 (1999)
arXiv:hep-th/9909062.
}
%
\lref\SenXM{
A.~Sen,
``Universality of the tachyon potential,''
JHEP {\bf 9912}, 027 (1999)
arXiv:hep-th/9911116.
}
%
\lref\KutasovQP{
D.~Kutasov, M.~Marino and G.~W.~Moore,
``Some exact results on tachyon condensation in string field theory,''
JHEP {\bf 0010}, 045 (2000)
arXiv:hep-th/0009148.
}
%
\lref\GerasimovZP{
A.~A.~Gerasimov and S.~L.~Shatashvili,
``On exact tachyon potential in open string field theory,''
JHEP {\bf 0010}, 034 (2000)
arXiv:hep-th/0009103.
}

\lref\reid{
M.~Reid,
``La correspondance de McKay,''
S\'eminaire Bourbaki, 52\`eme ann\'ee, novembre 1999, no. 867, 
to appear in Ast\'erisque 2000
arXiv:alg-geom/9911165. For further references see 
http://www.maths.warwick.ac.uk/$\scriptstyle\sim$miles/McKay/
}

\lref\ReidZY{
M.~Reid,
``McKay correspondence,''
arXiv:alg-geom/9702016.
}

\lref\MayrAS{
P.~Mayr,
``Phases of supersymmetric D-branes on Kaehler 
manifolds and the McKay  correspondence,''
JHEP {\bf 0101}, 018 (2001)
arXiv:hep-th/0010223.
}

\lref\AnselmiSM{
D.~Anselmi, M.~Billo, P.~Fre, L.~Girardello and A.~Zaffaroni,
``Ale Manifolds And Conformal Field Theories,''
Int.\ J.\ Mod.\ Phys.\ A {\bf 9}, 3007 (1994)
arXiv:hep-th/9304135.
}

\lref\BuscherQJ{
T.~H.~Buscher,
``Path Integral Derivation Of Quantum Duality In Nonlinear Sigma Models,''
Phys.\ Lett.\ B {\bf 201}, 466 (1988).
}

\lref\RocekPS{
M.~Rocek and E.~Verlinde,
``Duality, quotients, and currents,''
Nucl.\ Phys.\ B {\bf 373}, 630 (1992)
arXiv:hep-th/9110053.
}

\lref\DelaOssaXK{
X.~De la Ossa, B.~Florea and H.~Skarke,
``D-branes on noncompact Calabi-Yau manifolds: K-theory and monodromy,''
Nucl.\ Phys.\ B {\bf 644}, 170 (2002)
arXiv:hep-th/0104254.
}

\lref\ito{
Y.~Ito,
``Special McKay correspondence,''
arXiv:alg-geom/0111314.
}

\lref\morelli{
R. Morelli, ``K theory of a toric variety,'' Adv. in Math. {\bf 100}(1993)154
}

\lref\AspinwallXS{
P.~S.~Aspinwall and M.~R.~Plesser,
``D-branes, discrete torsion and the McKay correspondence,''
JHEP {\bf 0102}, 009 (2001)
arXiv:hep-th/0009042.
}
\lref\DiaconescuEC{
D.~E.~Diaconescu and M.~R.~Douglas,
``D-branes on stringy Calabi-Yau manifolds,''
arXiv:hep-th/0006224.
}
\lref\DiaconescuBR{
D.~E.~Diaconescu, M.~R.~Douglas and J.~Gomis,
``Fractional branes and wrapped branes,''
JHEP {\bf 9802}, 013 (1998)
arXiv:hep-th/9712230.
}

\lref\LercheVJ{
W.~Lerche, P.~Mayr and J.~Walcher,
``A new kind of McKay correspondence from non-Abelian gauge theories,''
arXiv:hep-th/0103114.
}

\lref\WittenCD{
E.~Witten,
``D-branes and K-theory,''
JHEP {\bf 9812}, 019 (1998)
arXiv:hep-th/9810188.
}
\lref\GarciaCompeanRG{
H.~Garcia-Compean,
``D-branes in orbifold singularities and equivariant K-theory,''
Nucl.\ Phys.\ B {\bf 557}, 480 (1999)
arXiv:hep-th/9812226.
}
\lref\DelaOssaXK{
X.~De la Ossa, B.~Florea and H.~Skarke,
``D-branes on noncompact Calabi-Yau manifolds: K-theory and monodromy,''
Nucl.\ Phys.\ B {\bf 644}, 170 (2002)
arXiv:hep-th/0104254.
}

\lref\GovindarajanVI{
S.~Govindarajan and T.~Jayaraman,
``D-branes, exceptional sheaves and quivers on 
Calabi-Yau manifolds: From Mukai to McKay,''
Nucl.\ Phys.\ B {\bf 600}, 457 (2001)
arXiv:hep-th/0010196.
}
\lref\GovindarajanEF{
S.~Govindarajan, T.~Jayaraman and T.~Sarkar,
``On D-branes from gauged linear sigma models,''
Nucl.\ Phys.\ B {\bf 593}, 155 (2001)
arXiv:hep-th/0007075.
}

\lref\HeCR{
Y.~H.~He,
``On algebraic singularities, finite graphs and D-brane gauge theories:
A  string theoretic perspective,''
arXiv:hep-th/0209230.
}

\lref\TakayanagiXT{
T.~Takayanagi,
``Tachyon condensation on orbifolds and McKay correspondence,''
Phys.\ Lett.\ B {\bf 519}, 137 (2001)
arXiv:hep-th/0106142.
}

\lref\TomasielloYM{
A.~Tomasiello,
``D-branes on Calabi-Yau manifolds and helices,''
JHEP {\bf 0102}, 008 (2001)
arXiv:hep-th/0010217.
}

\lref\GovindarajanVI{
S.~Govindarajan and T.~Jayaraman,
``D-branes, exceptional sheaves and quivers on Calabi-Yau manifolds:
{}From  Mukai to McKay,''
Nucl.\ Phys.\ B {\bf 600}, 457 (2001)
arXiv:hep-th/0010196.
}

\lref\BatyrevJU{
V.~V.~Batyrev and D.~I.~Dais,
``Strong Mckay Correspondence, String Theoretic Hodge Numbers And
Mirror Symmetry,''
arXiv:alg-geom/9410001.
}

\lref\ItoZX{
Y.~Ito and M.~Reid,
``The McKay correspondence for finite subgroups of SL(3,C),''
arXiv:alg-geom/9411010.
}

\lref\itonak{
Y.~Ito and H.~Nakajima,
``McKay correspondence and Hilbert schemes in dimension three,''
arXiv:al-geom/9802120.
}

\lref\crawthesis{
A.~Craw,
``The McKay correspondence and representations of the McKay quiver,''
Ph.D. thesis, University of Warwick; available at
http://www.math.utah.edu/~craw.
}

\lref\KachruAN{
S.~Kachru, S.~Katz, A.~E.~Lawrence and J.~McGreevy,
``Mirror symmetry for open strings,''
Phys.\ Rev.\ D {\bf 62}, 126005 (2000)
arXiv:hep-th/0006047.
}

\lref\GovindarajanEF{
S.~Govindarajan, T.~Jayaraman and T.~Sarkar,
``On D-branes from gauged linear sigma models,''
Nucl.\ Phys.\ B {\bf 593}, 155 (2001)
arXiv:hep-th/0007075.
}

\lref\HellermanCT{
S.~Hellerman and J.~McGreevy,
``Linear sigma model toolshed for D-brane physics,''
JHEP {\bf 0110}, 002 (2001)
arXiv:hep-th/0104100.
}

\lref\GovindarajanKR{
S.~Govindarajan and T.~Jayaraman,
``Boundary fermions, coherent sheaves and D-branes on Calabi-Yau
manifolds,''
Nucl.\ Phys.\ B {\bf 618}, 50 (2001)
arXiv:hep-th/0104126.
}

\lref\DistlerYM{
J.~Distler, H.~Jockers and H.~J.~Park,
``D-brane monodromies, derived categories and boundary linear sigma
models,''
arXiv:hep-th/0206242.
}

\lref\KatzGH{
S.~Katz and E.~Sharpe,
``D-branes, open string vertex operators, and Ext groups,''
arXiv:hep-th/0208104.
}

\lref\MorrisonYH{
D.~R.~Morrison and M.~Ronen Plesser,
``Towards mirror symmetry as duality 
for two dimensional abelian gauge  theories,''
Nucl.\ Phys.\ Proc.\ Suppl.\  {\bf 46}, 177 (1996)
arXiv:hep-th/9508107.
}

\lref\kronheimer{PB. Kronheimer and H. Nakajima, 
``Yang-Mills instantons on ALE gravitational 
instantons,'' Math. Ann. {\bf 288}(1990)263}

\lref\mooresegal{G. Moore and G. Segal, unpublished. The 
material is available in lecture notes from the April 2002 Clay 
School on Geometry and Physics, Newton Institute,  and 
http://online.kitp.ucsb.edu/online/mp01/moore1.} 

\lref\iaslectures{E. Witten, in {\it Quantum Fields and Strings: A 
Course for Mathematicians}, vol. 2, P. Deligne et. al. eds. 
Amer. Math. Soc. 1999}

\lref\kapustinlectures{A. Kapustin, Lectures at the KITP workshop 
on Mathematics and Physics, August, 2003.} 

\lref\SinYM{
S.~J.~Sin,
``Comments on the fate of unstable orbifolds,''
Phys.\ Lett.\ B {\bf 578}, 215 (2004)
[arXiv:hep-th/0308028].
}

\lref\LeeAR{
S.~g.~H.~Lee and S.~J.~H.~Sin,
``Chiral rings and GSO projection in orbifolds,''
Phys.\ Rev.\ D {\bf 69}, 026003 (2004)
[arXiv:hep-th/0308029].
}
\lref\LeeSS{
S.~Lee and S.~J.~Sin,
``Localized tachyon condensation and G-parity conservation,''
arXiv:hep-th/0312175.
}

\lref\GregoryYB{
R.~Gregory and J.~A.~Harvey,
``Spacetime decay of cones at strong coupling,''
Class.\ Quant.\ Grav.\  {\bf 20}, L231 (2003)
[arXiv:hep-th/0306146].
}

\lref\HeadrickYU{
M.~Headrick,
``Decay of C/Z(n): Exact supergravity solutions,''
arXiv:hep-th/0312213.
}

\lref\DixonIZ{
L.~J.~Dixon and J.~A.~Harvey,
``String Theories In Ten-Dimensions Without Space-Time Supersymmetry,''
Nucl.\ Phys.\ B {\bf 274}, 93 (1986).
}

\lref\MaldacenaKY{
J.~M.~Maldacena, G.~W.~Moore and N.~Seiberg,
``Geometrical interpretation of D-branes in gauged WZW models,''
JHEP {\bf 0107}, 046 (2001)
[arXiv:hep-th/0105038].
}

\lref\MinwallaHJ{
S.~Minwalla and T.~Takayanagi,
``Evolution of D-branes under closed string tachyon condensation,''
JHEP {\bf 0309}, 011 (2003)
[arXiv:hep-th/0307248].
}

\lref\KapustinRC{
A.~Kapustin and Y.~Li,
``D-branes in topological minimal models: The Landau-Ginzburg approach,''
arXiv:hep-th/0306001.
}
\lref\KapustinGA{
A.~Kapustin and Y.~Li,
``Topological correlators in Landau-Ginzburg models with boundaries,''
arXiv:hep-th/0305136.
}
\lref\KapustinKT{
A.~Kapustin and D.~Orlov,
``Lectures on mirror symmetry, derived categories, and D-branes,''
arXiv:math.ag/0308173.
}

\lref\DouglasSW{
M.~R.~Douglas and G.~W.~Moore,
``D-branes, Quivers, and ALE Instantons,''
arXiv:hep-th/9603167.
}

\lref\segal{G.B.  Segal, ``Equivariant K-Theory,''  Publ. Math.
IHES, {\bf 34}(1968)129}

\lref\MP{
  G.~W.~Moore and A.~Parnachev,
  ``Localized tachyons and the quantum McKay correspondence,''
  JHEP {\bf 0411}, 086 (2004)
  [arXiv:hep-th/0403016].
}

\lref\HV{
  K.~Hori and C.~Vafa,
  ``Mirror symmetry,''
  arXiv:hep-th/0002222.
}

\lref\HIV{
 K.~Hori and C.~Vafa,
  ``Mirror symmetry,''
  arXiv:hep-th/0002222;
 K.~Hori, A.~Iqbal and C.~Vafa,
  ``D-branes and mirror symmetry,''
  arXiv:hep-th/0005247;
}

\lref\boalch{Boalch, Oxford PhD thesis. Go to http://www.math.columbia.edu/~boalch/publications.html }

\lref\kaminski{D. Kaminski and R.B. Paris, ``Asymptotics of a class of multidimensional Laplace-type integrals. I. 
Double integrals,'' Phil. Trans. R. Soc. London. A (1998){\bf 356}583}

\lref\HV{Hori and Vafa, ``Proof'' } 
\lref\boalch{Boalch, Oxford PhD thesis. Go to http://www.math.columbia.edu/~boalch/publications.html } 

\lref\kaminski{D. Kaminski and R.B. Paris, ``Asymptotics of a class of multidimensional Laplace-type integrals. I. 
Double integrals,'' Phil. Trans. R. Soc. London. A (1998){\bf 356}583} 
 
\lref\HarveyNA{
  J.~A.~Harvey, D.~Kutasov and E.~J.~Martinec,
  ``On the relevance of tachyons,''
  arXiv:hep-th/0003101.
}

\lref\MelnikovHQ{
  I.~V.~Melnikov and M.~R.~Plesser,
  ``The Coulomb branch in gauged linear sigma models,''
  arXiv:hep-th/0501238; I.~V.~Melnikov and M.~R.~Plesser,
  ``A-Model Correlators from the Coulomb Branch,''
  arXiv:hep-th/0507187.
}

\lref\MNP{
  D.~R.~Morrison, K.~Narayan and M.~R.~Plesser,
  ``Localized tachyons in C(3)/Z(N),''
  JHEP {\bf 0408}, 047 (2004)
  [arXiv:hep-th/0406039].
}

\lref\MorrisonJA{
  D.~R.~Morrison and K.~Narayan,
  ``On tachyons, gauged linear sigma models, and flip transitions,''
  JHEP {\bf 0502}, 062 (2005)
  [arXiv:hep-th/0412337].
}

\lref\SarkarRY{
  T.~Sarkar,
  ``On localized tachyon condensation in C**2/Z(n) and C**3/Z(n),''
  Nucl.\ Phys.\ B {\bf 700}, 490 (2004)
  [arXiv:hep-th/0407070].
}

\lref\OoguriVR{
  H.~Ooguri, C.~Vafa and E.~P.~Verlinde,
  ``Hartle-Hawking wave-function for flux compactifications,''
  arXiv:hep-th/0502211.
}

\lref\McGreevyCI{
  J.~McGreevy and E.~Silverstein,
  ``The tachyon at the end of the universe,''
  arXiv:hep-th/0506130.
}

\lref\HorowitzVP{
  G.~T.~Horowitz,
  ``Tachyon condensation and black strings,''
  arXiv:hep-th/0506166.
}

\lref\HeadrickHZ{
  M.~Headrick, S.~Minwalla and T.~Takayanagi,
  ``Closed string tachyon condensation: An overview,''
  Class.\ Quant.\ Grav.\  {\bf 21}, S1539 (2004)
  [arXiv:hep-th/0405064].
}

\lref\dubrovinI{B. Dubrovin, ``Painleve' transcendents and 
two-dimensional topological field theory,''
arXiv:math.AG/9803107}
\lref\dubrovinII{B. Dubrovin, ``Geometry and analytic theory of 
Frobenius manifolds,'' arXiv:math\ .AG/9807034}
\lref\guzzetti{D. Guzzetti, ``Stokes matrices and monodromy of the 
quantum cohomology of
projective spaces,'' arXiv:math.AG/9904099}
\lref\ueda{K.Ueda, ``Stokes matrices for the quantum cohomologies of 
Grassmannians,''
arXiv:\ math.AG/0503355}
\lref\arouxI{D. Auroux, L. Katzarkov, and D. Orlov, ``Mirror symmetry 
for weighted projective
planes and their noncommutative deformations,'' arXiv:math.AG/0404281}
\lref\arouxII{D. Auroux, L. Katzarkov, and D. Orlov, ``Mirror symmetry 
for Del Pezzo surfaces:
vanishing cycles and coherent sheaves,'' arXiv:math.AG/0506166}

\lref\DouglasHQ{
M.~R.~Douglas and B.~Fiol,
``D-branes and discrete torsion. II,''
arXiv:hep-th/9903031.
}
\lref\BerkoozIS{
M.~Berkooz and M.~R.~Douglas,
``Five-branes in M(atrix) theory,''
Phys.\ Lett.\ B {\bf 395}, 196 (1997)
[arXiv:hep-th/9610236].
}
\lref\BerkoozKM{
M.~Berkooz, M.~R.~Douglas and R.~G.~Leigh,
``Branes intersecting at angles,''
Nucl.\ Phys.\ B {\bf 480}, 265 (1996)
[arXiv:hep-th/9606139].
}


\Title{\vbox{\baselineskip12pt
\hbox{hep-th/0507190}
}}
{\vbox{\centerline{Profiling the Brane Drain}
\vskip.06in
\centerline{in a Nonsupersymmetric Orbifold } 
}}
\centerline{Gregory Moore and Andrei Parnachev}
\bigskip
\centerline{{\it Department of Physics, Rutgers University}}
\centerline{\it Piscataway, NJ 08854-8019, USA}
 \vskip.1in \vskip.1in \centerline{\bf Abstract}  
\noindent
We study D-branes in a nonsupersymmetric orbifold of type $\IC^2/\Gamma$, perturbed 
by a tachyon condensate, using a gauged linear sigma model. The RG flow has both 
higgs and coulomb branches, and each branch supports different branes. 
The coulomb branch branes account for the ``brane drain'' from the higgs branch, 
but their precise relation to fractional branes has hitherto been unknown. 
Building on the results of hep-th/0403016 we construct, in detail, the map between 
fractional branes and the coulomb/higgs branch branes for two examples in the type 0 theory. 
 This map depends on the phase of the tachyon condensate 
in a surprising and intricate way. In the mirror 
Landau-Ginzburg picture the dependence on the tachyon phase is manifested by 
discontinuous changes in the shape of the D-brane.

\vfill

\Date{July 20, 2005}
   

\newsec{Introduction and summary}

An important property of string theory is that it is 
well-defined in the presence of certain spacetime 
singularities which render general relativity and quantum 
field theory ill-defined. Moreover, string theory 
contains mechanisms for smoothing out spacetime singularities. 
An interesting set of concrete examples  
of this phenomenon   are spacetime nonsupersymmetric orbifolds of flat space 
\AdamsSV. (See \MartinecTZ\HeadrickHZ\ for reviews).
In such situations, the closed string spectrum contains tachyons whose
wavefunctions are localized near the singular point; the resolution of singularities
happens through the condensation of these tachyons.
The presence of   $\NN=2$ worldsheet supersymmetry imposes
constraints on the dynamics of the system, allowing one to  
 understand the renormalization group (RG) flow \hkmm\ in a 
way analogous to the understanding of open string tachyon 
condensation  (see e.g. \HarveyNA). In this paper we assume that 
the RG flow gives a good description of  condensation
of localized closed string tachyons.
The behavior of the system in the IR corresponds 
to later times in the time evolution.

It is technically convenient to introduce the gauged linear 
sigma model (GLSM), whose higgs branch in the ultraviolet is the
nonsupersymmetric orbifold \VafaRA\mm.
In the process of RG flow, the higgs branch resolves into
a smooth \HJ\ space, which has a natural spacetime interpretation.
The number of branes wrapping nontrivial two-cycles
of the \HJ\ space is generally smaller than
that of fractional branes, so naively a ``brane drain'' is taking place.
However one must bear in mind that the infrared theory contains a coulomb branch 
with isolated massive vacua.
It has been suggested in \mm\ that D-branes wrapping  nontrivial
cycles in the higgs branch, together with D-branes supported 
at the vacua of the coulomb branch, are in one-to-one correspondence with 
fractional branes in the orbifold theory.
This picture has been sharpened in \MP\ where the open string
Witten index was used to construct a map between the fractional
branes and higgs and coulomb branch branes in the IR.\foot{
The significance of the coulomb branch goes back to 
\WittenYC\MorrisonFR\
and was also recently emphasized 
in \MelnikovHQ\ where the correlators in the topologically
twisted A-model were found to have support precisely on the
coulomb branch.}
While \MP\ found the map in the case of type II string theory,
a similar construction is possible in type 0 theory.
We describe it in detail in Section 2.

One can study D-branes away from the conformal point 
using techniques developed in \HIV.
The mirror description of the GLSM is given by the Landau-Ginzburg (LG)
theory with an inhomogeneous superpotential.
Fractional branes localized at the orbifold fixed point
preserve B-type supersymmetry, so they become A-type
branes in the LG model.
The latter are associated with critical points of the superpotential:
the critical points away from the origin give rise to the coulomb
branch branes, while those at the origin describe
the higgs branch brane(s).
In this paper we focus on the overlap of the boundary state
with the identity operator (this is defined more
precisely in Section 3).
This quantity, which we call the generalized central charge, should
in the first approximation be thought of as generalization of the D-brane mass to nonconformal
theories.
The generalized central charge 
of a brane described by the boundary state $|B\rangle$ can be computed as an integral
\eqn\gcci{  \langle B|1\rangle=\int\int {dx_1\over x_1} {dx_3\over x_3} \exp(-W)   }
over the A-brane surface.
In \gcci\ $x_1$ and $x_3$ are the LG fields, and $W$ is the superpotential. 
To determine the shape of the A-brane, it is necessary to
solve certain soliton equations.
The set of all solutions is parametrized by a small circle (wavefront) around
the critical point.
The A-brane surface is traced by the wavefront evolving in 
time \HIV.

We specialize to the case of $\IC^2/\IZ_{n(p)},\,p=1$ orbifolds, 
whose minimal resolution contains a single non-trivial cycle.
(The higgs branch in this case is $\OO(-n)\ra\IP^1$.)
There is a single higgs branch brane wrapping the nontrivial cycle,
and $n-2$ coulomb branch branes associated with the massive vacua of the
superpotential.
We find that the generic form of the A-brane surface resembles
that of a propeller.
Near the critical point, the wavefront is a small circle, whose segments
at late times trace various quarter-planes (``wings'' of the ``propeller'').
This property can be used to compute \gcci\ as a function of (complex) tachyon
expectation value $w$; $w\ra 0$ corresponds to the orbifold (UV) limit,
while $|w|\ra\infty$ describes the IR regime.
As explained in Section 4, \gcci\ satisfies a GKZ equation, which 
in the Calabi-Yau case is a Picard-Fuchs equation for the periods.
A basis of nonconstant solutions of the GKZ equation
is given by the integrals \gcci\ over the quarter planes.
Linear combinations of these integrals, which we compute in Section 4, determine
the generalized central charge of the A-brane whose wings asymptote to these quarter planes.

We analyze the behavior of \gcci\ in the simple cases of $n=3,4$ in Section 5.
The intersection matrix and the quantum symmetry of the orbifold theory 
suffices to determine the map between the coulomb branch branes
and the fractional branes.
This map depends of the phase of the tachyon expectation value.
Multiplication of $w$ by an $n$-th root of unity enforces the permutation
symmetry of the fractional branes.
In terms of the propeller surfaces, the asymptotics change
discontinuously when the value of $w$ goes from one angular sector 
of the complex plane to another.

As explained in Section 2, even after modding out by permutation symmetry
of the fractional branes, in the type 0 theory there is more then one expression for the
higgs branch brane which is consistent with the intersection matrix.
In section 5 we find that each of the $n$ angular sectors which differ by a 
permutation of fractional branes is further divided into smaller
subsectors, where different expressions for the higgs branch brane are realized.
The generalized central charge for all higgs branch branes has the same leading
logarithmic behavior in the regime of large $|w|$.

To summarize, this paper is organized as follows.
In the next section we describe the orbifold, GLSM and
its mirror LG model.
We use the intersection matrix to relate
fractional branes to the coulomb and higgs branch branes in the IR.
In Section 3 we define the generalized central charge.
In Section 4 we show that it solves the GKZ equation, and analyze
the solutions.
Section 5 is devoted to the detailed analysis of the $n=3$ and
$n=4$ cases.
We discuss our results and directions for future research in
Section 6.
Appendix A contains information on the construction of fractional
brane boundary states.
Appendix B is devoted to the numerical analysis of the shape
of A-branes in the LG theory.

\newsec{Fractional branes vs. higgs and coulomb branch branes } 
In this section we start by reviewing the results of  \mm and \MP\
where the fate of fractional branes was studied using the gauged
linear sigma model.
We recall the mirror description in terms of the Landau-Ginzburg theory,
and give a first hint at the appearance of the fractional branes
in this language.
We study the intersection form for type II and type 0 orbifolds
and use it to construct the linear map between the fractional
branes and the LG branes.
We discuss the $\IC^2/\IZ_{3(1)}$ and  $\IC^2/\IZ_{4(1)}$
examples in detail.
Part of this section is review material.
A more detailed exposition can be found in \hkmm\mm\MP\MartinecTZ.
The map between the fractional branes and the LG (or, equivalently, GLSM)
branes is spelled out in detail, although such a map was
constructed implicitly in \MP.
The discussion of the intersection form in the type 0 theory is new.
The intersection form in type 0 theory is important for the $\IC^2/\IZ_{3(1)}$
example which is discussed at the end of this section and later in the paper. 
\subsec{Condensation of localized tachyons and the fate of the fractional branes.}

We consider type II or type 0 theory in 9+1 dimensions.
The orbifolding by $\IZ_{n(p)}$ happens in the 67 and 89 planes,
parametrized by complex coordinates $X^{(1)}$ and  $X^{(2)}$.
The orbifold group is generated by
\eqn\orbaction{ g=\exp\left({2\pi i\over n}(J_{67}+p J_{89})\right) ,}
where $J_{67}$ and $J_{89}$ generate rotations in two complex planes.
When there are fermions in the theory, $p$ is defined ${\rm mod} \, 2n $.
We will take the fundamental domain to be $p\in (-n,n)$.
The action of $g^n$ on the Ramond sector ground state is a multiplication by $(-1)^{p\pm 1}$,
depending on chirality.
When $p$ is even, this acts as $(-1)^F$ where $F$ is the spacetime fermion
number.
In type II, there is no bulk tachyon and there are closed string
fermions in the bulk, hence $p$ must be odd \refs\AdamsSV.
In the NSR formalism, a useful ingredient in the theory is the operator \hkmm
\eqn\defxj{
  X^{(i)}_{s\over n}=\sigma_{s/n}^{(i)}\;\exp\left[i(s/n)(H^{(i)}-\tilde H^{(i)})\right];
	\qquad i=1,2; \qquad s=1,2,\cdots, n-1
}
where $H^{(i)},\tilde H^{(i)}$ are the bosonised left- and right-moving fermions and
$\sigma_{s/n}$ is the bosonic twist $s$ operator \refs\DixonQV.
In the following we will restrict our attention to the left movers.
There are two possible inequivalent choices for
the worldsheet $\NN=1$ supersymmetry generator in the theory.
Correspondingly, there are two chiral rings
which are BPS under these supersymmetries.
The (c,c) ring vertex operators are
\eqn\xj{X_{s\over n}^{(cc)}=X^{(1)}_{s\over n} X^{(2)}_{\left\{ {sp\over n} \right\}}       ,}
where $\left\{ x \right\} \equiv x-\left[ x \right]$
is the fractional part of $x$.
The (c,a) ring vertex operators are
\eqn\xjca{X_{s\over n}^{(ac)}=
          X^{(1)}_{s\over n} \left( X^{(2)}_{1-\left\{ {sp\over n} \right\}}\right)^*       ,}
The operation $p\ra -p$ corresponds to exchanging the $(c,c)$ and the $(c,a)$
rings. 
Therefore we can restrict ourselves to the theories with $p\in (0,n)$.

The generators of the (c,a) ring, denoted  $W_\alpha, \;\;\alpha=1\ldots r$ 
form a collection of (in general) relevant operators. Turning these on 
in the action induces RG flow to  
the minimal resolution of the singularity \refs{\hkmm,\mm}.
For the $\IC^2/\IZ_{n(p)}$ orbifold such a resolution is encoded
in the continued fraction
\eqn\cfdef{{n\over n-p}=a_1-{1\over a_2-{1\over a_3- \ldots}}\; := \; [a_1,a_2,\ldots a_r]      ,}
Note the appearance of $n-p$ rather then $p$ in \cfdef, since we are
talking about the (c,a) ring.
The smooth space which appears after the minimal
resolution of the singularity is called the \HJ\ manifold.
The generators of the chiral ring are in one-to-one correspondence
with the exceptional $\IP^1$'s of this space.
Their intersection numbers are given by
\eqn\exc{C_{\alpha\beta}=
         -\delta_{\alpha,\beta-1}+a_\alpha \delta_{\alpha,\beta}-\delta_{\alpha,\beta+1}}
The (c,c) ring generators give rise to the resolution with 
similar properties; one needs to substitute $n-p\ra p$ in \cfdef.
In the type II theory one should bear in mind the existence of a chiral GSO projection.
As explained in \MP, all generators $W_\alpha$ in the (c,a) ring
survive the GSO projection if and only if all  $a_j$ are even integers.
In the (c,c) ring at least one generator is projected out \MP.
In the type 0 theory chiral operators in (c,a) and (c,c) 
rings are not projected out by the diagonal GSO projection.
Since $p\ra p+n$ does not affect the ring structure, but only
affects the GSO projection, the closed string sector in type 0 theories  
is unchanged under this operation.
This means that type 0 theories with $p$ and $n-p$ are isomorphic: they are related
by interchanging the (c,c) and the (c,a) rings.
Put differently, in a type 0 theory with a given $p$, one should
be able to resolve the singularity into two different \HJ\ spaces,
whose intersection numbers correspond to continued
fractions determined by both $p$ and $n-p$.
In \MP\ it has been shown that branes in type II theory which wrap nontrivial two-cycles 
in the \HJ\ space are given by linear combinations of the
fractional branes.
In this paper we will see that the situation in the type 0
string theory is similar.
Since there are two different resolutions, one can define two
sets of branes wrapping the two-cycles in these spaces.
Both sets are given by certain linear combinations of fractional branes.

In \mm\ the gauged linear sigma model (GLSM) was used to shed light
on the fate of fractional D-branes in the process of twisted
tachyon condensation.
The field content of the relevant GLSM involves
$r$ abelian $\CN=2$ gauge fields
$V_\alpha$, $\alpha=1,...,r$ coupled to $r+d$ 
$\CN=2$ chiral matter fields
$X_i$ with charges 
\eqn\qcharges{ Q_{\alpha i} = -C_{\alpha i}    }
with $C_{\alpha i}$ given by [compare with \exc]:
\eqn\exci{C_{\alpha i}=
         -\delta_{\alpha,i-1}+a_\alpha \delta_{\alpha,i}-\delta_{\alpha,i+1};\qquad i=0,\ldots,r+1}
The field strengths of the gauge fields 
are contained in twisted chiral superfields
$\Sigma={1\over 2} \{\overline\CD,\CD^*\}$.
The classical Lagrangian is
\eqn\glsmact{
  \CL = \int d^4\theta \;\left(\Xbar_i e^{2Q_{\alpha i} V_\alpha} X_i
	-\frac{1}{2e_{\alpha}^2}\bar\Sigma_\alpha \Sigma_\alpha\right)
	-\half\left(\int d^2\tilde\theta 
		\;t_{\alpha}\Sigma_\alpha+{\rm c.c.}\right)\ ,
}
where repeated indices are summed and 
\eqn\teebare{
t_{\alpha } =\zeta_\alpha-i\theta_\alpha
}
 combines
the Fayet-Iliopoulos (FI) parameter $\zeta$ and theta angle $\theta$
for the $\alpha^{\rm th}$ gauge field; $d^2\tilde\theta$ is
the twisted chiral superspace measure.  
The renormalized FI parameters at the scale $\mu$ is
\eqn\FIren{
  t_{\alpha,{\rm eff}}(\mu) = t_{\alpha,{\rm bare}} 
	+ \sum_{i=1}^{r+d} Q_{\alpha i} \log{\mu\over \Lambda} 
}
where  $t_{\alpha,{\rm bare}} $ are bare parameters 
defined at the momentum cutoff scale $\Lambda$. 
Due to \qcharges, \exc,\  $t_{\alpha,{\rm bare}}\ra-\infty $
in the UV.
As explained in \mm,\ in this regime the higgs branch
describes the $\IC^2/\IZ_{n(p)}$ orbifold.
The theory also contains a coulomb branch, where the lowest
components of $\Sigma_\alpha$ get expectation values.
In the infrared, the higgs branch becomes the \HJ\ space \mm.
We will call branes wrapping two-cycles in this resolved space ``higgs
branch branes''.
Another important property of the infrared physics is
decoupling of the coulomb branch from the higgs branch.
The former develops a set of massive isolated minima.
Some fractional branes become B-branes ``supported'' at these coulomb 
branch vacua, as twisted tachyons condense \mm.
We call such branes ``coulomb branch branes''.

For our purposes it will be more convenient to look at the ``mirror''
description of the coulomb branch that follows from the 
approach to mirror symmetry using abelian duality of 2D gauge theory 
of Morrison and Plesser
\MorrisonFR\MorrisonYH. This can 
be cast in terms of an effective Landau-Ginzburg theory \HIV. 
(We follow the line of argument explained in \mm.)
The chiral superfields $X^i$ are eliminated in favor of 
the twisted scalar superfields $Y_i$.
The twisted superpotential in the theory reads
\eqn\horivafa{\eqalign{
        \widetilde W&= \sum_{\alpha=1}^{r} \Sigma_\alpha
\Biggl(\sum_{i=0}^{r+1} Q_{\alpha i} nY_i-t_\alpha(\mu)\Biggr)
                        +\mu \sum_i \lambda_i e^{-nY_i}\cr}
}
%
%
%
Integrating out the $Y_i$ gives the effective superpotential
of \WittenYC\MorrisonFR, while
eliminating instead the $\Sigma_\alpha$ and $Y_\alpha$ 
$\alpha=1,...,r$, gives
(in terms of $u_{0}=(\mu\lambda_0)^{1/n}\exp[-Y_{0}]$ 
and $u_{r+1}=(\mu\lambda_{r+1})^{1/n}\exp[-Y_{r+1}]$)
\eqn\twspotl{
  \widetilde W = u_0^n + u_{r+1}^n 
		+ \sum_{\alpha=1}^r
                \lambda_\alpha' u_0^{p_\alpha}u_{r+1}^{q_\alpha}\ ,
}
where 
\eqn\newlambds{
  \lambda_{\alpha}' = \lambda_{\alpha} 
	\;\Lambda^{1-\Delta_{\alpha}}
	\;e^{t'_{\alpha,{\rm bare}}} 
	= \lambda_{\alpha}
		\;\mu^{1-\Delta_{\alpha}}
		\;e^{t'_{\alpha,\eff}(\mu)}
\ .
}
%
The scaling dimensions of the $\alpha$'s operator in the sum
identifies it with the ring generator $W_\alpha$.
Two important comments are in order

\item{(1)} The LG description involves dualizing
the phases of $X_i$, and hence is not well defined 
near the higgs branch, where $X_\alpha=0$.
We do expect it to give a correct description of the
coulomb branch though.
As we will see later in the paper, we will be able to 
recover the description of the higgs branch as well.
The essential ingredient will be the identification of
the higgs branch brane with a combination of the fractional
branes with the help of intersection form.
\item{(2)}
The `mirror' $\IZ_n$ transformation
\eqn\mirzn{
  (u_0,u_{r+1})\sim(\omega u_0,\omega^{-p}u_{r+1})
}
leaves the effective superpotential \twspotl\ invariant --
it fixes all the $\Sigma'_\alpha$.  Indeed it is a 
{\it gauge symmetry} remnant of the duality transformation 
and therefore we should quotient the LG model by its action.  

\noindent
In this paper we will be concerned with the simplest 
case of a continued fraction of length one.
This corresponds to $p=1$, $r=1$, $a_1=n$.
There are two fields, $x_1\equiv u_0$ and $x_3\equiv u_2$.
The tachyon expectation value is determined by the
parameter $w\equiv \lambda'_1$.
The LG superpotential is
\eqn\lgsp{ W=x_1^n+w x_1 x_3+x_3^n  }
and the theory should be quotiented by 
\eqn\lgq{ (x_1,x_3)\sim (e^{2\pi i\over n}x_1, e^{-{2\pi i\over n}}x_3) }

%
%
%
%
\subsec{Intersection form and the map between LG and fractional branes}
We start by reviewing the results of \MP.
There we considered the D-brane intersection form in type II string theory
on the orbifold $\IC^2/\Gamma$ \refs{\DouglasHQ\BerkoozIS-\BerkoozKM}
\eqn\defind{\I_{ab}=\tr_{R,ab} (-1)^F q^{L_0-{c\over 24}}   .}
Here the trace is over the states of the open string suspended
between D-branes which correspond to representations of $\Gamma$
labeled by $a$ and $b$ and $F$ is the worldsheet fermion
number.
In the case of type II theory, this formula can be
written as \MP
\eqn\om{\I_{ab}={4 \over n}\sum_{s=0}^{n-1}\exp\left({2\pi i (b-a)s \over n}\right)
                           \sin\left({\pi s\over n}\right)\sin\left({\pi s p\over n}\right) }
The first factor in the sum comes from the action of the group
element on the Chan-Phaton factors, while the product of the $sin$'s 
is due to the fermion zero modes in the R sector.
It is not hard to evaluate \om: 
\eqn\oma{\I_{ab}=\delta_{a-b-{1-p\over 2}}+\delta_{a-b+{1-p\over 2}}
           -\delta_{a-b-{1+p\over 2}}-\delta_{a-b+{1+p\over 2}}         }
where
\eqn\defdm{\delta_a\equiv\delta_{a, 0\;\mn}.}
Note that the arguments of delta functions in \oma\ are
always integers, thanks to the requirement that $p$ is odd.
The matrix $\I$ in \oma\ is written in the basis 
\eqn\bbasic{ e_0,e_1,\ldots,e_{n-1}   }
where $e_a$ is the $a$-th fractional brane\foot{The details of fractional
brane construction in type 0 and type II string theories are summarized
in Appendix A.}.
The intersection matrix $\I$ is invariant under the cyclic
permutation of fractional branes, $e_a\ra e_{a+1}$.
In \MP\ we found a change of basis which block-diagonalizes
$\I$; one of the two blocks is given precisely by the intersection
matrix for the higgs branch branes $C_{\alpha\beta}$.
However this block-diagonalization is clearly invariant under
the cyclic permutation of the fractional branes; in other
words, if certain expressions for the higgs branch branes
\eqn\hone{ h_\alpha=\sum_a H_{\alpha a} e_a, \quad H_{\alpha a}\in \IZ  } 
give rise to the intersection matrix $C_{\alpha\beta}$, then
the expressions with the indices of fractional branes permuted,
\eqn\htwo{ h_\alpha=\sum_a H_{\alpha a} e_{a+1}  } 
are equally good.
We will see later in the paper that, thanks to the
dependence of the map between fractional and coulomb/higgs branch branes
on the tachyon VEV, all possible cyclic permutation 
of fractional branes are realized, depending on the phase of the
tachyon expectation value.
For now we assume the basis \bbasic\ for simplicity, keeping the permutation symmetry in 
mind.
In fact, it is convenient to switch to the basis 
$\sum_a e_a,e_1,\ldots,e_{n-1}$ which effectively substitutes both the
first row and the first column in $\I$ by a set of zeroes 
(since the D0 brane, $D0=\sum_a e_a$, has zero intersection with any brane
in the theory, including itself).
Now we can omit the first row and column from $\I$-- it
is this reduced matrix, denoted $\It$ which appears in the rest of this paper, unless
stated otherwise.

We would like to generalize this discussion for type 0.
The number of branes is now doubled, for there are
branes which are labeled by the choice of sign 
in the gluing conditions (see Appendix A for details).
We will call these branes $\eta=+1$ or $\eta=-1$ branes\foot{
These branes are often called electric and magnetic branes
in the literature, but they are not electric/magnetic duals.
In type 0 theories obtained as orbifolds by $(-1)^{F_{\rm spacetime}}$ 
of the type II string, they are in fact fractional branes \MP.}.
As explained in \MP, when $p$ is odd, the intersection form
of the type 0 theory is simply obtained from the intersection
form $\It$ of type II:
\eqn\izero{ \It_0=\pmatrix{
~0& ~\It\cr
~\It^T& ~0 }
}
where $\It^T=\It$. 
Note that fermionic degrees of freedom only exist on the intersections
of branes of different types.
It will be convenient to separate $\eta=+1$ and  $\eta=-1$
fractional branes into
two sets: the ones that are labeled by the ``special representation'' integer $e_{p_\alpha}$
and the rest, $e_\nu$.
The integers $p_\alpha$ are determined by the continued fraction $[a_1,a_2,\ldots a_r]$
via the recursion relations \mm:
\eqn\rrpq{
p_{j-1}/p_j=\left[a_j,a_{j+1},\ldots,a_r\right],\qquad 1\le j\le r
}
with the initial conditions $p_{r+1}=0,\;p_r=1$.
In the basis 
\eqn\basise{  e^{(+)}_{p_1},\ldots,e^{(+)}_{p_r},\{e^{(+)}_{\nu}\},
                     e^{(-)}_{p_1},\ldots,e^{(-)}_{p_r},\{e^{(-)}_{\nu}\}         }
we consider the following ansatz for the linear map between
the LG branes (higgs and coulomb branch branes $h_\alpha$  and $c_\alpha$)
and the fractional branes
\eqn\az{ \pmatrix{~h^{(+)}_\alpha\cr~c^{(+)}_\alpha\cr~h^{(-)}_\alpha\cr~c^{(-)}_\alpha}=
\pmatrix{
~A& ~0\cr
~0& ~B } \pmatrix{~e^{(+)}\cr e^{(-)}} = 
\pmatrix{~1& ~a& ~0& ~0\cr ~0& ~1& ~0& ~0\cr ~0& ~0& ~1& ~b^T\cr ~0& ~0& ~0& ~1}
\pmatrix{ ~e^{(+)}_{p_\alpha}\cr ~e^{(+)}_{\nu}\cr ~ ~e^{(-)}_{p_\alpha}\cr ~e^{(-)}_{\nu}  }
} 
where $a$ and $b$ are matrices whose entries are integers.
We then require
\eqn\cb{  \pmatrix{~A& ~0\cr ~0& ~B } \pmatrix{~0& ~\It\cr ~\It^T& ~0 } \pmatrix{~A^T& ~0\cr ~0& ~B^T }=
   \pmatrix{~0& ~0& ~C& ~0\cr ~0& ~0& ~0& ~C'\cr ~C& ~0& ~0& ~0\cr ~0& ~C'& ~0& ~0}     }   
One can solve for $a$, $b$ and $C$.
The result is 
\eqn\abc{   a=-x C'^{-1},\qquad b=-C'^{-1} y,\qquad C=C_1-x C'^{-1} y    }
where $C_1$, $C'$, $x$ and $y$ are the components of $\It$:
\eqn\icomps{   \It=\pmatrix{~C_1& ~x\cr ~y& ~C' }          }
Note that the map determined by \abc\  
is determined up to an addition/subtraction of any multiple of D0 branes,
since the latter have zero intersection with any fractional brane. 

When $p$ is odd, the orbifold group is $\IZ_n$, one can define type II theory,
and $\It=\It^T$.
In this case $\It$ is given by the reduction of \oma, and $a=b^T$.
In \MP\ it was shown that if in addition all $a_\alpha$ are even, then
the entries of $a$ (and $b$) in \abc\ are integers 
and $C$ computed in \abc\ coincides with \exc.
The closed string sector of type 0 theory is invariant 
under $p\ra p+n$.
However eq. \om\ is not invariant under such a shift.
This is a manifestation of the fact that one can define two sets
of branes in the type 0 theory.
These two types of branes will have intersection matrices
corresponding to two different types of resolution, as discussed above.
A simple set of examples considered in \MP\ is $p=1, n-1$.
The intersection form for the $p=1$ case reproduces the Cartan matrix 
which defines the supersymmetric ALE singularity.
For $p=n-1$ there is a single higgs branch brane $h$ which wraps the base of $\OO(-n)\ra \IP^1$
which is the \HJ\ manifold in this case.
The change of basis in the $p=n-1$ case is nontrivial; here we quote 
the result obtained in \MP:
\eqn\cbpn{   h=e_1+2 e_2+\ldots+{n\over 2}e_{n\over 2}-
    \left({n\over 2}-1\right)e_{{n\over 2}+1}-\ldots-e_{n-1}           }

Suppose now $p$ is even.
(And, consequently, $n$ is odd as we restrict our 
consideration to $n$, $p$ relatively prime.)
In this case the orbifold group is $\IZ_{2n}$ and expression \om\
for the unreduced matrix $\I_0$ should be modified to
\eqn\omb{\I_{0,ab}={4 \over 2 n}\sum_{s=0}^{2n-1}\exp\left({2\pi i (a-b)s \over 2 n}\right)
                           \sin\left({\pi s\over n}\right)\sin\left({\pi s p\over n}\right) }
The indices $a$, $b$ which label the branes now run from $0$ to $2n-1$.
$n$ is substituted by $2n$ in the prefactor and in the first factor
in the sum-- this is the result of the order of the orbifold
group becoming $\IZ_{2n}$.
The matrix element in \omb\ evaluates to zero whenever $a-b$ is even.
This is an indication that $a$, $b$ now label not only different types
of fractional branes, but also $\eta=+1$ and $\eta=-1$ types.
We can define $a=2 a'$ and $b=2b'+1$ with $a'=0,\ldots,n-1$  and  $b'=0,\ldots,n-1$ 
labeling $\eta=+1$ and $\eta=-1$ branes respectively.
The intersection form is then of the form \izero, although 
$\I$ is no longer a symmetric matrix:
\eqn\inonsym{  \I_{a'b'}=\delta_{a'-b'+{p\over 2}}+\delta_{a'-b'+1-{p\over 2}}
           -\delta_{a'-b'-{p\over 2}}-\delta_{a'-b'+1+{p\over 2}}         }
To use \abc\ we still need to factor out the $D0$ brane by omitting
the first row and the first column.
At this point it is worth mentioning the following
important issue that was not present in the type II theory.
We can shift the $\eta=-1$ branes by $e^{(-)}_{a'}\ra e^{(-)}_{a'+1}$,
without shifting the  $\eta=+1$ branes.
This leads to permutation of columns, $\I_{a',b'}\ra \I_{a',b'+1}$.
After the reduction to the subspace which does not contain a D0 brane,
we obtain $n$ inequivalent matrices $\It$ this way.
In this paper we mostly discuss the $p=1$ case for $n=3,4$.
As explained in Section 5, all inequivalent matrices $\It$ obtained as
described above, admit a block-diagonalization of the form \cb.
The analysis for general $n$ and $p$ will be reported elsewhere.


\newsec{D-branes in the LG mirror and generalized central charges}

The superpotential of the LG model providing a mirror description of the GLSM \glsmact\
has a set of critical points.
According to \HIV\ each critical point gives rise to a
D-brane with A-type boundary conditions.
This D-brane, which we will sometimes refer to as an A-brane surface,
is a Lagrangian submanifold of $\IC^2$ whose
image in the $W$-plane is a half-line which starts at the critical value and extends
in the positive real direction.
The critical points away from $(x_1,x_3)=(0,0)$ correspond
to the coulomb branch branes, while the single critical point
at the origin gives rise to a higgs branch brane
(recall that we specialize to the case of continued fraction of
length one)

As explained in \HIV\ a practical way to determine 
these surfaces is to consider the soliton equation
\eqn\soleq{  {d x^i\over d\sigma}={1\over2} g^{i\bar j} \p_{\bar j} \bar W   }
where $ g_{i\bar j}$ is the Kahler metric and $\sigma$ is the coordinate
along the soliton trajectory.
The soliton trajectory is supposed to originate from a critical point,
where the Kahler metric is nonsingular.
(This leads to problems when considering the higgs branch brane, whose
critical point is at $(x_1,x_3)=(0,0)$ where the Kahler metric
cannot even be reliably determined).
The shape of the A-brane associated with a given critical point is
the set of all trajectories satisfying \soleq.
Near the critical point this shape can be easily determined.
One needs to find the coordinates $u^i$ which diagonalize the system \soleq.
In these coordinates, the solutions are $u^i=u^i_0 \exp(\lambda^i \sigma)$,
so that $u^i\ra0$ as $\sigma\ra-\infty$.
Consistency of the equations \soleq\ then forces $u^i_0 \in \IR$.
A convenient way of parameterizing a set of solutions  \HIV\ is considering
a sphere of small radius $\epsilon$
\eqn\sphere{  \sum_i (u^i)^2=W(\phi^i_*+u^i)-W(\phi^i_*)=\epsilon^2,\qquad u^i\in \IR       }
As $\sigma$ increases, this small sphere (``the wavefront'') will evolve; the surface traced by it
in this process is the A-brane.
In Appendix B we analyze these surfaces numerically.
We find that in the models we consider they resemble a propeller,
with various segments of the small circle developing into quarter-plane ``wings''.

An interesting object that one can consider in the LG models
is an overlap of a RR ground state\foot{
The correspondence is realized by performing the worldsheet path integral
on the semiinfinitely long cigar with no insertions, but in the twisted theory,
to produce a RR state.},
corresponding to the identity operator,
with the D-brane specified by the boundary state $| B\rangle$ \HIV.
To make this object holomorphic, one needs to consider
the formal limit $\bar W\ra 0$ \HIV.
In this case, the overlap can be computed as an integral over the
A-brane surface:
\eqn\gcc{  \langle B|1\rangle=\int\int {dx_1\over x_1} {dx_3\over x_3} \exp(-W)   }
As we will see in the next section, 
this integral is convergent, since $W\ra +\infty$ in the asymptotic region.
Moreover, in the next section we will also see that \gcc\ 
satisfies the GKZ equation.
In the supersymmetric case the solutions 
would give rise, via mirror symmetry, to the integrals
of the complexified Kahler form over the cycles in the
higgs branch of the GLSM.
This motivates us to call the quantity \gcc\ ``the generalized 
central charge'' of the D-brane described by the boundary state $|B\rangle$.
In the non-Calabi-Yau case of this paper,
the higgs branch brane in the GLSM still has a geometric interpretation
of a brane wrapping a two-cycle in the \HJ\ space. 

Note that the branes described above correspond to the B-branes in the original GLSM and orbifold
theories with $\eta=1$.
Before performing the GSO projection, the theory contains a second set of
branes with $\eta=-1$, which preserve a different combination 
of the worldsheet $\NN=2$ supersymmetry.
Such branes are described by eq. \soleq\ with an additional
minus sign in the right-hand side.
The image of these branes in the $W$-plane would therefore be half-lines extending in the
negative real direction.
The definition \gcc\ would have to be modified accordingly.


\newsec{Generalized periods  for the $\IC^2/\IZ_{n(1)}$ orbifold }

In this section we consider the generalized periods for the  LG superpotential 
\eqn\lgpot{
W := a_1 x_1^n + a_2 x_1 x_3 + a_3 x_3^n 
}
This section is rather technical. A summary is found in the final subsection.

\subsec{GKZ Equation}

Associated to a toric manifold is a canonically determined system of differential 
equations, the GKZ system of differential equations. For toric hypersurfaces in 
CY manifolds these equations are related to the Picard-Fuchs equations of the mirror 
and therefore solutions are related to the periods of the mirror variety. 

In the present case, using the toric data of the $\IC^2/\IZ_{n(1)}$ manifold one
finds the differential equation 
\eqn\gkzeqi{
\biggl[ \Theta^2 -  z (n \Theta) (n \Theta +1) \cdots (n\Theta + n-1)\biggr] F(z)=0 
}
with $\Theta = z{ d\over dz}$. 
$z=\infty$ is a regular singular point, and corresponds to the orbifold point. 
$z=0$ is an irregular singular point and corresponds to the ``IR limit.''

To derive \gkzeqi\  one begins 
with the fan  $v_1 = (0,1), v_2 = (1,0), v_3 = (n,-1)$ to produce the differential 
operator 
\eqn\diffop{ 
\CD:= (\p_{a_1} \p_{a_3} -  (-1)^n\p_{a_2}^n)    
}
Now, again using the toric vectors one defines an invariant combination   $ z := a_1 a_3/a_2^n$.  
When acting on a function depending on $a_i$ only through $z$ it is 
straightforward to show that 
\eqn\simpdep{
 \CD f(z) = {1\over a_1 a_3} \biggl[ \Theta^2 -  z (n \Theta) (n \Theta +1) \cdots (n\Theta + n-1)\biggr] F(z)
}
This establishes \gkzeqi. 

A useful change of variable $w= z^{-1/n}=a_2/(a_1 a_3)^{1/n}$ brings the GKZ equation to the form: 
\eqn\gkzsimp{
\biggl[ -\bigl(-{d\over dw} \bigr)^{n} + {1\over n^2} \bigl(w{d\over dw}\bigr)^2 \biggr]
  f(w) =0 
}

The $n$-dimensional space of solutions to this equation is the $n$-dimensional space of 
generalized periods. The constant solution - which will be associated with the $D0$ brane - 
is somewhat trivial and we define $\CV$ to be the space of nonconstant solutions which vanish at 
$w=0$. Much of the work in   Section 5 of this paper will be 
writing down different bases for $\CV$  and interpreting them physically.  
One basis of solutions is obtained by straightforward application of the Frobenius technique. 
This is: 
\eqn\fpp{
\eqalign{
\hat f_{-1}(w)&:=1 \cr
\hat f_m(w) & := (-w)^{m+1} \sum_{k=0}^\infty (-w)^{nk} {\bigl(\Gamma(k + {m+1\over n})\bigr)^2\over \Gamma(kn+m+2) } 
\qquad 0 \leq m \leq n-2\cr}
}
Note that $\hat f_m(w)$ are entire functions in the $w$-plane for $n>2$ with an essential singularity at $w=\infty$. 
It will be useful to extend the definition \fpp\ to include $\hat f_{n-1}(w)$, although this function does 
{\it not} solve the GKZ equation.

\subsec{Integral representation} 

In the physical interpretation of the solutions it is very useful to have an integral 
representation directly related to the path integral which computes the overlap between 
RR groundstates and the branes. This solution has the schematic form 
\eqn\lgi{
\int_{\gamma} e^{- W}  {dx_1\over x_1} {dx_3 \over x_3}
}
where $\gamma$ is an appropriate contour, to be discussed in detail below.

Note that it is trivially true that for $W$ given by \lgpot\ and $\CD$ given by \diffop\
\eqn\lgiact{
\CD \int_{\gamma} e^{- W}  {dx_1\over x_1} {dx_3 \over x_3} =0
}
%
%
%
where $\gamma$ is any {\it fixed} chain of real dimension 2 in $\IC^* \times \IC^*$.

Now, the generalized periods should only be functions of the scaling variable $w$. 
This can be arranged by exploiting the $\IC^*\times \IC^*$ action on $(x_1,x_3)$. 
In fact, by analyticity, we need only consider $\IR^*_+ \times \IR^*_+$ invariant orbits. Any such orbit through $(\alpha,\beta) \in \IC^*\times \IC^*$ is of the 
form 
\eqn\gami{
\gamma_{\alpha,\beta} := \{ (t_1 \a, t_3 \b)\vert t_1,t_3>0 \}.
}
One might be tempted to use the chains $\gamma_{\alpha,\beta}$ in \lgi\ to produce solutions of 
\gkzsimp. There are two problems with this. 
 First,  we must also regularize the logarithmic singularities 
at the origin. Next,  we must 
ensure convergence of the integral at $\infty$. The singularity at the origin is easily regularized 
by taking   $t_i\geq \e$. Thus we consider  the contour integral:
\eqn\regin{
I_{\a,\b}(w;\e):= \int_\e^\infty {dt_1\over t_1} \int_\e^\infty {dt_3\over t_3} e^{-W} 
}
where  
\eqn\neve{
W = \a^n t_1^n + \a \b w t_1 t_3 + \b^n t_3^n
}
with $t_1, t_3>0$ always.

We will use this basic integral to construct solutions. To  ensure convergence at infinity  we require
$\re(\a^n)>0$ and $\re(\b^n)>0$. There are different sectors of the $(x_1,x_3)$ plane in which the 
integrals converge. We will need a way of denoting these {\it convergent sectors}.  
Define  $\a = \vert \a\vert e^{i \theta}$. Then if   $\a$ is in a convergent sector 
there must exist an integer  $s_\a$ such  that 
\eqn\voncen{
- {\pi \over 2n} <  \theta + {2\pi \over n} s_\a < {\pi \over 2n} 
}
We will denote the convergent sectors \voncen\ in the $\alpha$ plane by $\CS_{s_\a}$. 
Equation \voncen\ only defines $s_\a$ modulo $n$. If we choose the principal branch of the 
logarithm then $s_\a$ is defined absolutely. We choose the fundamental domains:
\foot{ This has one awkward feature for $n$ even: In this case 
the angular sector containing the negative real axis is a convergent sector and $s$ is discontinuous 
across the negative real axis, jumping from $s=-n/2$ just above the negative real axis to $s=n/2$
just below the negative real axis. }
\eqn\rangeess{
\eqalign{ 
& s= {n-1\over 2}, {n-3\over 2}, \dots, - {n-1\over 2} \qquad \qquad n \quad odd \cr
& s= \half n ,   \dots, - \half n \qquad \qquad n \quad even \cr}
}
See Figs. 1,2 for $n=3,4$. 
\midinsert\bigskip{\vbox{{\epsfxsize=3in
        \nobreak
    \centerline{\epsfbox{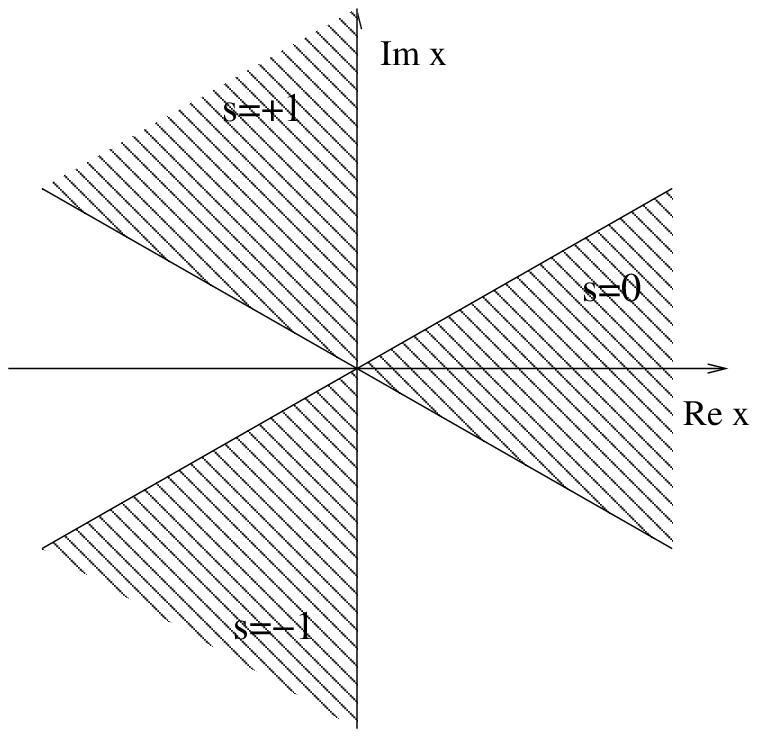}}
        \nobreak\bigskip
    {\raggedright\it \vbox{
{\bf Fig 1.}
{\it The angular structure in the $\alpha$, $\beta$ 
(equivalently, $x_1$, $x_3$) planes for $n=3$.  In the shaded sectors
the integral \regin\ converges.}}}}}}
\bigskip\endinsert
\midinsert\bigskip{\vbox{{\epsfxsize=3in
        \nobreak
    \centerline{\epsfbox{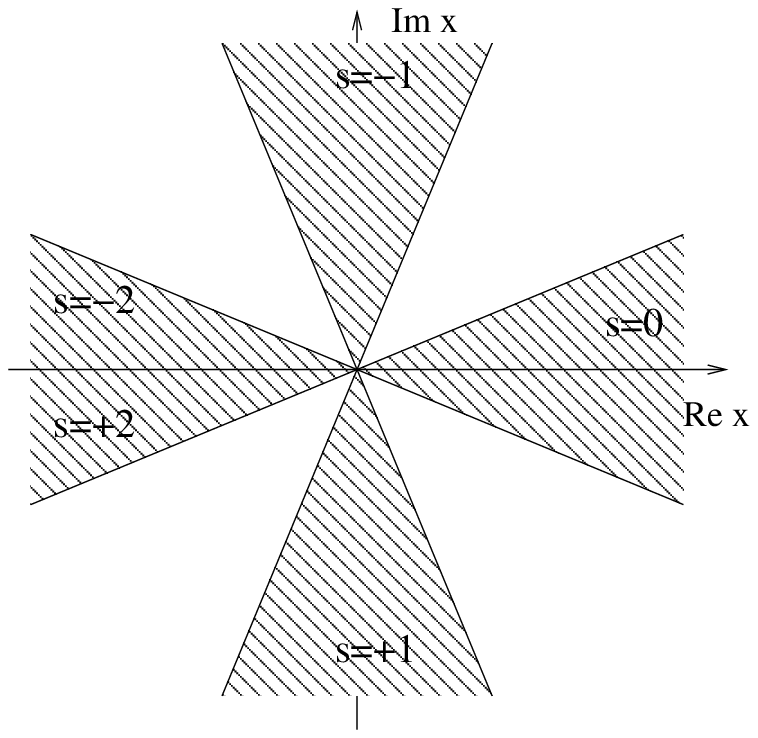}}
        \nobreak\bigskip
    {\raggedright\it \vbox{
{\bf Fig 2.}
{\it The angular structure in the $\alpha$, $\beta$ 
(equivalently, $x_1$, $x_3$) planes for $n=4$.  In the shaded sectors
the integral \regin\ converges.}}}}}}
\bigskip\endinsert

By using rescalings of $t_1,t_3$, the equation $\CD e^{-W}=0$, and  keeping track of boundary terms, and 
\simpdep\  it is straightforward to show that 
\eqn\difsnac{
a_1 a_3 \CD I_{\a,\b}(w;\e) =  {1\over n^2} e^{-W(t_1=\e, t_3 = \e)} 
- {\a^n\over n} \int_\e^\infty dt_1 t_1^{n-1} e^{-W(t_1, t_3=\e) }
 - {\b^n\over n} \int_\e^\infty dt_3 t_3^{n-1} e^{-W(t_1=\e, t_3) }
}
Thus, the $I_{\a,\b}(w;\e)$ do {\it not} solve the differential equation. 
Note, however, that the RHS of \difsnac\ has a smooth $\e\to 0$ limit, 
given by $-1/n^2$
%
%

We will form solutions from the $I_{\a,\b}$ by taking appropriate linear combinations
of $I_{\a,\b}(w;\e)$ and taking the $\e\ra0$ limit. 
In order to produce appropriate combinations let us investigate the $\epsilon \to 0$ 
behavior.  Note that $I_{\a,\b}(w;\e)$ has a   convergent expansion in $w$.  
If we expand the integrand
in \regin\  in a power 
series in $w$, only the first term has a divergence as $\e \to 0$. Let us define 
$\hat I_{\a, \b}(w;\e)$ via:
\eqn\ddfihat{
I_{\a, \b}(w;\e) =  \int_\e^\infty {dt_1\over t_1}e^{-\a^n t_1^n}  \int_\e^\infty {dt_3\over t_3} e^{-\b^n t_3^n}  + 
\hat I_{\a, \b}(w;\e)
}
The first term is divergent for $\epsilon \to 0$, and constant in $w$. 
In fact, one has: 
\eqn\consttrm{
\int_\e^\infty {dt_1\over t_1}e^{-\a^n t_1^n} = {1\over n} E_1(\epsilon^n \a^n) = - \log \e - {1\over n} \log \a^n - {\gamma\over n} + \CO(\e^n).
}
Thus we conclude that  
\eqn\ilime{
I_{\a,\b}(w;\e) \rightarrow (\log \e)^2  + K_{\a\b} \log \e  + {1\over n^2}(\log \a^n + \gamma)(\log \b^n + \gamma) + \hat I_{\a\b}(w;0) + \CO(\epsilon)
}%
In particular 
\eqn\lieim{
{d\over dw} I_{\a,\b}(w;\e)={d\over dw} \hat I_{\a,\b}(w;\e)
}
has a smooth $\e\to 0$ limit.

It follows from \ilime\ that if we choose linear combinations 
\eqn\linec{
\sum c_{\a\b} \hat I_{\a,\b}(w;0) 
}
such that 
\eqn\linde{
\sum c_{\a\b} =0  
}
then we have a solution of the GKZ equation \gkzsimp. Indeed, we can take the limit $\e\to 0$ 
directly from \ddfihat\   and 
we  can compute the power series in 
$w$ explicitly. Using
\eqn\simplin{
\int_0^\infty {dt\over t} t^k e^{-\a^n t^n} = \a^{-k} e^{-2\pi i {k\over n} s_\a} {1\over n} \Gamma({k\over n}) \qquad k\geq 1
}
  we arrive at 
\eqn\itoeff{
\eqalign{
\hat I_{\a,\b}(w;0) & = {1\over n^2} \sum_{k=1}^\infty (-w)^k {(\Gamma(k/n))^2\over k!} e^{-2\pi i {k\over n}(s_\a + s_\b)} \cr
 & = {1\over n^2} \sum_{j=1}^n e^{-2\pi i {j\over n} (s_\a + s_\b)} \hat f_{j-1}(w) \cr}
}
where $\hat f_m$ was defined in \fpp. 
Recall that $\hat f_{n-1}$ does {\it not} solve the differential equation \gkzsimp. This is in accord with \difsnac.

Let us note a few important properties of the functions $\hat I_{\a,\b}(w;0)$. First, it is clear from 
\itoeff\ that  $\hat I_{\a,\b}(w;0) $ only depends on $\a,\b$ through the sector $\CS_{s_\a} \times \CS_{s_\b}$.  
In fact $\hat I_{\a\b}$ only 
depends on the combination $s_{\a\b} := (s_\a + s_\b) \mod~ n$. By definition, the function  $\hat I_s$ is $\hat I_{\a\b} $ 
such that $s_{\a\b}=s \mod ~n$. Next, note that the functions
\eqn\newslkc{
\sum c_{s} \hat I_{s}(w;0)
}
 subject to $\sum c_{s}=0 $
span the space $\cal V$ of nonconstant solutions to the GKZ system.  
Indeed, inverting the finite Fourier transform we see that the span of $\hat I_{\a,\b}(w;0)$ 
is the span of $\hat f_0, \dots, \hat f_{n-1}$, and that $\hat f_m$ for $m<n-1$ are given by linear combinations 
with $\sum c_{\a\b}=0 $.  Finally, note   that $\sum c_{\a\b} I_{\a\b}(w;\epsilon)$ only has a smooth limit for 
$\sum c_{\a\b}=0$ and $\sum c_{\a\b} K_{\a\b}=0$. In this case 
$\lim_{\e\to 0} \sum c_{\a\b} I_{\a\b} = \sum c_{\a\b} \hat I_{\a\b} $. 

\subsec{LG symmetry}

The ``LG symmetry'' or ``quantum symmetry'' plays a  fundamental role in what follows. 
The point $w=0$ corresponds to the orbifold point, where there is a quantum symmetry that 
permutes the fractional branes. From the action of this symmetry on the chiral ring 
generator we see that the action of the  quantum symmetry is $w \to \omega w$, where $\omega  = e^{2\pi i /n}$, so a 
generator of the quantum symmetry takes $\hat I_s \to \hat I_{s+1}$, that is: 
\eqn\lgsymm{
\hat I_s(\omega^t w) = \hat I_{s-t}(  w)
}

Thus the set of solutions  $\hat I_1-\hat I_0, \hat I_2-\hat I_1, \dots, \hat I_{n-1}-\hat I_{n-2}, \hat I_0 - \hat I_{n-1}$ 
are cyclically permuted under the quantum symmetry. The sum of these solutions is $0$. Thus, this basis is 
reminiscent of the   the space of fractional branes orthogonal to the $D0$ brane, and one might 
be tempted to identify these with the periods of fractional branes.   Unfortunately, quantum monodromy is not 
strong enough to guarantee this and we will see that in fact a more subtle basis corresponds to the 
basis of fractional branes.

\subsec{Critical points and propeller branes}

The integral representation \lgi\ is useful for investigating the asymptotics of the solutions
via stationary phase.  
In this section we set $a_1=a_3=1$ so 
$$
W = x_1^n + w x_1 x_3 + x_3^n.
$$
Of course, solving for $W'=0$ is the same as solving 
for the LG vacua. We find there are $(n-1)^2$ solutions. $n(n-2)$ solutions are nonzero and come in 
$(n-2)$ different ``Landau-Ginzburg 
orbits'' of the quantum $\IZ_n$ symmetry $(x_1,x_3) \to (\omega x_1, \omega^{-1} x_3)$. 

We should simply write: 
\eqn\lgiiip{
\eqalign{ 
x_1 & = (-{w\over n})^{1\over n-2} e^{2\pi i  {\nu (n-1)\over n(n-2)} } \cr
x_3 & = (-{w\over n})^{1\over n-2} e^{2\pi i   {\nu \over n(n-2)}  } \cr}
}
with $\nu = 1, \dots, n(n-2)$.
%
%
The LG $\IZ_n$ symmetry is $\nu \to \nu + (n-2)$. 
The remaining critical point is at $(x_1,x_3)=(0,0)$.

The critical value of $W$ at \lgiiip\ is 
\eqn\lgivp{
W_\nu = (2-n) \bigl({-w\over n}\bigr)^{{n\over n-2}} e^{2\pi i {\nu  \over n-2}} 
}
 Note that the different LG orbits are  separated by the value of $W$ on the orbit. 
It is also useful to compute the Hessian: 

\eqn\hess{
W_\nu'' = - w\pmatrix{(n-1)e^{-2\pi i \nu/n} & - 1 \cr -1 & (n-1)e^{2\pi i \nu/n}\cr} 
}
 Note that 
$$
\det W'' = w^2 n(n-2)
$$
does not depend on the sector. All critical points are Morse critical points.

Associated to each critical point is a vacuum state in the GLSM. Associated with 
each vacuum is a (topological) D-brane.   To write the generalized period for the brane we 
define $\Gamma_\nu$ to be the A-brane surface
defined in \HIV (see Section 3). In the math 
literature these are known as  ``Lefshetz thimbles'' and in the context of this paper they 
are the propeller branes. We then define
\eqn\coulbm{
C_\nu(w):= \int_{\Gamma_\nu} e^{-W} {dx_1\over x_1}{dx_3\over x_3} 
}
These are well-defined for the nonzero critical points. 
The period $C_0$ for brane corresponding to $(x_1,x_3)=(0,0)$ cannot be obtained from 
\coulbm\ because the latter has a logarithmic 
singularity.

Although the chains $\Gamma_\nu$ are not $\IR^* \times \IR^*$ invariant the functions 
$C_\nu$ nevertheless satisfy the GKZ equation. The reason is that the chains $\Gamma_{\nu}$ 
approach a linear combination of the chains $\gamma_{\a\b}$ at infinity, and, by Cauchy's theorem 
we can deform contours in a compact region without changing the integral. In fact, by 
examining the asymptotic behavior of the functions $C_{\nu}(w)$ at large $w$ 
we find that $C_\nu$ with $\nu=0,1,\dots, n-2$  define a basis for the space $\CV$, and therefore there is a locally constant 
matrix $M_{\nu s}$ such that 
\eqn\matrx{
C_\nu = \sum_{s} M_{\nu s} \hat I_s 
}
This matrix is only locally constant. It will be constant in angular sectors and will change 
discontinuously across angular sectors because the solutions to an equation with an irregular singular 
point exhibit Stokes' phenomenon.

\subsec{A  basis of asymptotic solutions}

Let us now consider the asymptotic behavior of the functions in $\CV$ at large $w$.
From the saddle point formula the contribution of the critical point of type $\nu$ to $C_\nu$ is: 
\eqn\spcontr{
\epsilon_{\nu}  {1\over \sqrt{n^3(n-2)}} \bigl({-w\over n} \bigr)^{-n\over n-2} e^{-2\pi i {\nu\over n-2} } 
\exp\biggl[ (n-2) (-{w\over n})^{n\over n-2} e^{2\pi i {\nu\over n-2}} 
\biggr] 
}
where $\epsilon_\nu = \pm 1$. This is the leading term in an asymptotic  
expansion given by expanding the integrand of 
\eqn\asympsolii{
 \eqalign{
&  e^{-W_\nu} \int_{-\infty}^{+\infty} {dt_1 dt_3 \over (x_1  +t_1)
(x_3  +t_3)} e^{-\half \pmatrix{t_1 & t_3\cr}W''\pmatrix{t_1\cr t_3\cr} } \cr
& 
e^{- \sum_{j=3}^n {n\choose j} x_1^{n-j} t_1^j + \sum_{j=3}^n {n\choose j} x_3^{n-j} t_3^j }\cr}
}
in powers of $t_1, t_3$ and doing the gaussian integrals. 
Here $(x_1,x_3)$ is the critical point we are expanding around. 
Note that the ``interaction vertices'' in the Feynman diagram expansion depend on $\nu$. 
The result is that \spcontr\ is multiplied by   a power series in $w^{-n/(n-2)}$.

These asymptotic expansions   give $n-2$ asymptotic solutions 
to the GKZ equation, valid at large $w$. The last solution is the log solution from the 
asymptotic expansion at $(0,0)$. 
Thus, we have a basis of formal solutions: 
\eqn\asympbasis{
\eqalign{
b_0 & =  \log w - \sum_{k=1}^\infty {\Gamma(kn)\over (k!)^2} w^{-kn}  \cr 
b_\nu & = {1\over \sqrt{n^3(n-2)}} \bigl({-w\over n} \bigr)^{-n\over n-2} e^{-2\pi i {\nu\over n-2} } 
\exp\biggl[ (n-2) (-{w\over n})^{n\over n-2} e^{2\pi i {\nu\over n-2}} 
\biggr] 
p^{\nu}  \qquad \nu = 1, \dots, n-2 \cr}
}
where $p^\nu$ is an asymptotic  series in $w^{-n/(n-2)}$.

Note that, roughly speaking, half of the series $b_\nu$ are ``exponentially growing,'' and 
half are ``exponentially decreasing.'' Thus, there is a filtration on $\CV$ given by the maximal 
asymptotic growth as $w\to \infty$ along a fixed ray. We order the values of $\nu_i$ so that 
\eqn\filrta{
\re (-{w\over n})^{n\over n-2} e^{2\pi i {\nu_1\over n-2}} < \re (-{w\over n})^{n\over n-2} e^{2\pi i {\nu_2\over n-2}}< \cdots 
}
to produce the filtration 
\eqn\filtrate{
\CF_1 \subset \CF_2 \subset \cdots \subset \CF_{n-1} = \CV 
}
Note that there are $(n-1)$ steps in the filtration because of the logarithmic solution.

Our next task is to find out how the true solutions $\sum_s c_s \hat I_s(w)$ with $\sum_s c_s=0$ 
fit into the filtration 
\filtrate.

\subsec{Asymptotics of  $\hat I_s$: exponential growth }

The most natural way to investigate the asymptotics of $\hat I_s$ is to apply the 
saddle-point technique to the integral representation
\eqn\inthat{{d\over dw } \hat I_{\alpha\beta}=-\int_{\gamma_{\alpha\beta}}e^{-W}dx_1dx_3 .}
Applying the saddle point technique is not straightforward. Care is needed in 
establishing   which of the saddle points \lgiiip\ contribute to a 
given integral $\hat I_s$. However, if a critical point lies in the convergent sector 
containing the contour defining $\hat I_s$ then that critical point does contribute since 
no large contour deformations are required. Using that rule alone we can learn some useful 
facts about when $\hat I_s(w)$ has exponential growth.

Let us label the convergent sectors by $\CS_s$ where $s$ is defined, modulo $n$, by \voncen.

Depending on the phase of $w$ the critical points $(x_1,x_3)_\nu$ in 
\lgiiip\ might or might not be in a convergent sector. Note that since $x_1^\nu= x_3^\nu e^{2\pi i {\nu\over n}}$, 
it follows that 
$x_1,x_3$ either both lie in a convergent or in a nonconvergent sector. 
The condition to lie in a convergent sector is:  
\eqn\secdep{
\re\biggl(
(-{w\over n})^{n\over n-2} e^{2\pi i \bigl( {\nu \over n-2} \bigr) } \biggr) > 0 
}
This is precisely the criterion 
that $W_\nu <0$, and hence the same as the criterion that the 
contribution to an integral, if it exists, is always a {\it growing} 
exponential.

Let us analyze more fully when the exponentially growing 
critical points can contribute to an integral $I_{\a\b}$. 
Let  
\eqn\ampphs{
(-{w\over n})^{1\over n-2} e^{2\pi i \bigl( {\nu \over n(n-2)} \bigr) } = e^{i \psi }A 
}
where $A>0$, and $-\pi < \psi < \pi$. Then from \secdep\ we know there is an $s_*$ with 
\eqn\vonceni{
- {\pi \over 2n} <  \psi + {2\pi \over n} s_* < {\pi \over 2n} 
}
Then the critical point \lgiiip\ is 
in the sector: 
\eqn\insectr{
(A e^{i \psi + 2\pi i\nu /n} , A e^{i \psi } ) \in \CS_{s_*+\nu} \times \CS_{ s_* }
}
Thus, for $w$ such that \vonceni\ holds, $\hat I_{\a\b}$ has an exponential growth from a critical point iff 
$s_{\a\b} = 2s_* +\nu  \mod ~n $. Moreover, the Landau-Ginzburg symmetry $\nu \to \nu+(n-2)$ relates the 
growth of different functions $\hat I_s \to \hat I_{s+1}$. To exploit this consider  
$\hat I_0(w)$. For $\vert \arg(-w)^{1/(n-2)} \vert < {\pi\over 2}$, that is, for 
\eqn\expgr{
-{\pi \over 2} + {\pi\over n} < \arg(-w) < {\pi \over 2} - {\pi\over n}
}
the critical point $\nu=n(n-2) \cong 0$ contributes to the integral. Therefore, $\hat I_0$ has exponential 
growth in this sector. Now using $\hat I_s(w) = \hat I_0(\omega^{-s} w)$ we can make similar statements 
about the other sectors.

As we shall see in the next section, in the sector complementary to \expgr\ $\hat I_0$ has in fact at most 
logarithmic growth. In overlapping sectors of the type \expgr\ we can form linear combinations of the 
$\hat I_s$ to produce functions with exponential growth slower than the leading one.

{\bf Examples}

\item{1.} $n=3$. The functions $\hat I_s$ have exponential growth 
like $\CE=-\sqrt{27}w^{-3}\exp((-w/3)^3)$ for: 
\eqn\expnsg{
\eqalign{
\hat I_0: \quad & \quad   -{\pi\over 6} < \arg(-w) < {\pi \over 6} \cr
\hat I_1: \quad & \quad   -{5\pi \over 6} < \arg(-w) < - {3\pi \over 6} \cr
\hat I_2: \quad & \quad   {3\pi \over 6} < \arg(-w) < {5\pi \over 6}\cr}}
These are the shaded regions in Fig. 3 below.

\item{2.} $n=4$. We have exponential growth for 
\eqn\expnsg{
\eqalign{
\hat I_0: \quad & \quad   -{\pi\over 4} < \arg(-w) < {\pi \over 4} \cr
\hat I_1: \quad & \quad   {\pi\over 4} < \arg(-w) < {3\pi \over 4} \cr
\hat I_2: \quad & \quad   {3\pi\over 4} < \arg(-w) < {5\pi \over 4} \cr
\hat I_3: \quad & \quad   {-3\pi\over 4} < \arg(-w) < {-\pi \over 4} \cr}
}
See Fig. 4.

\subsec{Asymptotics of $\hat I_s$: Coefficient of the logarithm}

In this section we introduce a different integral representation for the functions $\hat I_s$ 
which, while only valid in part of the complex $w$ plane, is very useful for extracting 
asymptotic behavior for $\vert w \vert \to \infty$. We apply a method described in  
 \kaminski. \foot{
This reference examines asymptotics of the integrals relevant to 
the more general set of  $\IC^2/\IZ_{n(p)}$ models and might be useful in further 
extensions of the present paper. }

In what follows $\arg(z)$ always means the principal branch of the logarithm, so it is defined 
for $z\in \IC - \IR^-$ and $\vert \arg(z)\vert < \pi$. 
We begin with 
\eqn\ezee{
e^{-z} = {1\over 2\pi i } \int_{\CC} \Gamma(u) z^{-u} du \qquad \vert \arg(z)\vert < \pi/2
}
where $\CC$ is the contour $u= \epsilon + i y , y\in \IR, \epsilon>0$. Note that 
\eqn\gamms{
\vert \Gamma(x+ iy ) \vert \sim \sqrt{2\pi} \vert y\vert^{-\half + x} e^{-{\pi \over 2} \vert y \vert} 
}
for $y \to \pm \infty$ at any fixed $x$. Thus the integrand converges absolutely for 
$\vert \arg(z)\vert < \pi/2$. 
To prove \ezee\ note that we can close the contour in 
the left half plane.

We apply this to $e^{-\a\b w t_1 t_3}$ in the integral representation 
for $\hat I$ which follows from \regin. Thus we must require 
\eqn\newrpi{
\vert \arg(\a\b w) \vert < {\pi \over 2} 
}
and for such values of $\a,\b, w$ we define the integer $N(\a,\b, w)$ by 
\eqn\newrpii{
-{\pi \over 2} <  \arg(\a) + \arg(\b) + \arg(w) + 2\pi N(\a,\b, w) < {\pi \over 2} 
}
The integrals are absolutely convergent and we can exchange them and do the $t_1, t_3$ 
integral. Using the generalization of \simplin\
\eqn\simplinb{
\int_0^\infty {dt\over t} t^{-s} e^{-\a^n t^n} = \a^{s-1} e^{ {2\pi i s_\a\over n} (s-1)} {1\over n} \Gamma({1-s\over n}) \qquad \re(s)<1 
}
 we find
\eqn\newrep{
{d\over dw} \hat I_{\alpha\beta}(w;0) = - e^{-2\pi i s_{\a\b}/n} {1\over 2\pi i n^2} 
\int_{\CC} \Gamma(u) \bigl(\Gamma({1-u\over n})\bigr)^2 w^{-u} 
\exp\bigl[ 2\pi i u \bigl({s_{\a\b}\over n} - N(\a,\b, w) \bigr) \bigr] du
}
This is valid in the region \newrpi. It is a good exercise to use \gamms\ to check that \newrep\ is 
an absolutely convergent integral in this range. Indeed, this  
condition guarantees absolute convergence of the integral in \newrep\ along 
any contour of the form $x + i y$ for fixed $x$ with $y\in \IR$.

Now, using the property that $I_{\a\b}$ only depends on the wedge in which $\a,\b$ live 
we can map out the range of validity for the integral representation \newrep. 
It follows that, for $\hat I_0$, the domain of validity of \newrep\ is  the region 
\eqn\logreg{
-{\pi \over 2} - {\pi \over n} < \arg(w) < {\pi \over 2} + {\pi \over n} 
}
Note that this is perfectly complementary to the
region \expgr.
Using the LG symmetry we find that the formula \newrep\ for  $\hat I_s$ 
holds in the     range
\eqn\rane{
- {\pi \over 2 } + {\pi \over n}(2s-1) < \arg w <  {\pi \over 2 } + {\pi \over n}(2s+1) 
}
%
If $n$ is sufficiently larger then $s$ then $N=0$ throughout \rane.

The integral \newrep\ is useful because it allows us to obtain asymptotic expansions for $\hat I_s$ 
in the region \rane.  While we   cannot close the $u$-contour integral in the  
left half plane,
we can displace the contour to the right, thanks to \gamms. In the process we pick up poles 
from $u=1+kn$, $k=0,1,2,\dots$.  In this way we arrive at the   asymptotic expansion 
\eqn\asympttos{
{d\over dw} \hat I_{\alpha\beta}(w;0)  \sim 2\pi i ({s_{\a\b}\over n} - N(\a,\b,w) ) \sum_{k=0}^\infty {\Gamma(1+kn)\over (k!)^2} 
w^{-1-kn}  + \sum_{k=0}^\infty D_k(n) w^{-1-kn}
  }
where the $D_k(n)$ are 
functions of $k,n$ but are independent of $w$ and, crucially, are 
independent  of $\a,\b$ and hence 
cancel out when one forms combinations $\sum c_{\a\b} \hat I_{\a\b}$ 
such that $\sum c_{\a\b}=0$. Integrating this formula we have the asymptotic expansion
\eqn\asymptosi{
\eqalign{
\hat I_{\a\b}(w;0) & \sim 2\pi i ({s_{\a\b}\over n} - N(\a,\b,w) )\biggl( \log w - \sum_{k\geq 1} {\Gamma(kn)\over (k!)^2} w^{-kn} \biggr)  + \CU \cr
\CU & := - \half(\log w)^2 - {(n-2)\over n}\gamma \log w + c+ \sum_{k\geq 1} {\Gamma(kn)\over (k!)^2} (-kn w^{-kn }\log w + h_k w^{-kn}) \cr
h_k & := -1 + {2\over n}(1+ {1\over 2} + \cdots + {1\over k} - \gamma) - \Psi(1+kn) \cr}
}
where $c$ is a constant. 
The important thing in this formula is that $\CU$ is 
independent of $\alpha,\beta$ (and hence of $s$). 
Note that the $s$-dependent term is nicely consistent with 
the logarithmic dependence on $w$ in the expansion $\CU$,
and the LG symmetry.
The formula is valid in the LG images of \logreg. 
The asymptotics  perfectly complement the region \expgr\ where the leading exponential dominates.

One important conclusion we can draw from \asymptosi\ is that  
in those regions where \newrpi\ is simultaneously valid for all terms with 
$c_{\a\b}\not=0$ and such that 
$\sum  c_{\a\b} ({s_{\a\b}\over n} - N(\a,\b,w) ) =0 $
then $\sum c_{\a\b} \hat I_{\a\b}$ will be an exponentially decaying solution.

\def\im{{\rm Im}}

The discussion in this section falls short of giving a complete description of the filtration \filtrate\ 
in all angular sectors for general $n$ because we have not explained how to form linear combinations with 
prescribed sub-exponential growth.  One can apply the saddle point technique to the integral \newrep\ 
for linear combinations, such as $\hat I_s + \hat I_{-s} - 2\hat I_0$ for which the pole terms cancel. 
One finds that $\hat I_s$ contributes an exponential behavior like 
\eqn\expgrwth{
 \cases{ \exp\biggl[ - (n-2) \bigl({w\over n}\bigr)^{n\over n-2}   e^{-{2\pi i \over n-2}(s+1) }\biggr] & $\im w>0 $ \cr
\exp\biggl[ - (n-2) \bigl({w\over n}\bigr)^{n\over n-2}   e^{-{2\pi i \over n-2}(s-1) }\biggr] & $\im w<0 $ \cr}
}
in the intersection of the regions \rane. (We assume $n \gg \vert s\vert $ at this point.) 
The term with the least rapid decay will then dominate.
Careful application of this rule might  suffice to determine the
full filtration \filtrate\ but we have not carried this out.
There is also a   Mellin-Barnes representation of 
the functions $\hat f_m$ but the saddle point technique applied to this representation  proves 
inconclusive.  Nevertheless, the results we have presented here do suffice to give a rather complete picture of the 
filtration for $n=3$ and $n=4$, as we describe in the next subsection.

\subsec{Examples: $n=3$ and $n=4$}

Let us show how the above general results can give a   complete picture of the filtration 
for the cases $n=3$ and $n=4$. Let 
\eqn\calee{
\CE = -\sqrt{27} w^{-3} \exp\bigl(-{w\over 3}\bigr)^3
}
Then we have
\eqn\cyou{
\hat I_0 \sim \cases{ \CU & $-{5\pi \over 6} < \arg w < {5\pi \over 6} $\cr
 \CE & $-{\pi \over 6} < \arg -w < {\pi \over 6} $\cr}
}
\eqn\cyou{
\hat I_1 \sim \cases{ \CU + {2\pi i \over 3} \log w + \cdots & $-{\pi \over 6} < \arg w < \pi  $\cr
\CU - {4\pi i \over 3} \log w + \cdots & $-\pi < \arg w < -{\pi \over 2}  $\cr
 \CE & $-{\pi \over 2} < \arg -w < -{\pi \over 6} $\cr}
}
\eqn\cyou{
\hat I_{-1} \sim \cases{ \CU + {4\pi i \over 3} \log w + \cdots & ${\pi \over 2} < \arg w < \pi  $\cr
\CU - {2\pi i \over 3} \log w + \cdots & $-\pi < \arg w < {\pi \over 6}  $\cr
 \CE & ${\pi \over 6} < \arg -w < {\pi \over 2} $\cr}
}

Now let us describe the corresponding filtrations. 
There are only two steps in \filtrate. In the ``convergent sectors'' 
$\re w^3 <0 $ the exponential solution, which is asymptotic to $\CE$, 
 is growing and we have $\CF_0 \subset \CF_+$ where $\CF_0$ is 
the 1-dimensional space spanned by the log solution. In the sectors with $\re w^3>0$, $\CE$ is decaying
and we have the filtration $\CF_- \subset \CF_0$, where $\CF_-$ is the one-dimensional 
space spanned by the exponentially decaying solution $\CE$. 

The first step in the filtration is given by 
\eqn\firsta{
\CF_0 = {\rm Span}\cases{\hat I_0 - \hat I_{-1} \sim {2\pi i \over 3} \log w + \cdots & $-{3\pi \over 6} < \arg w < - {\pi \over 6} $\cr
\hat I_1 - \hat I_{-1} \sim - {2\pi i \over 3} \log w + \cdots & $-{\pi \over 6} < \arg (-w) <  {\pi \over 6} $\cr
\hat I_0 - \hat I_{1} \sim -{2\pi i \over 3} \log w + \cdots & ${\pi \over 6} < \arg w <  {3\pi \over 6} $\cr}
}
for the sectors in which $\CE$ is exponentially growing. Note the three lines of \firsta\ are related by LG symmetry. 
Similarly the first step in the filtration is given by  
\eqn\firstap{
\CF_- = {\rm Span}\cases{2\hat I_0 - \hat I_{-1}- \hat I_{+1}  \sim cnst. \CE & $-{\pi \over 6} < \arg w < {\pi \over 6} $\cr
2\hat I_1 - \hat I_{-1}- \hat I_{0}  \sim cnst. \CE & ${3\pi \over 6} < \arg w < {5\pi \over 6} $\cr
2\hat I_{-1} - \hat I_{1}- \hat I_{0}  \sim cnst. \CE & $-{5\pi \over 6} < \arg w < -{3\pi \over 6} $\cr}
}
for the sectors in which $\CE$ is exponentially decreasing.

Similarly, for $n=4$ we find: 
\eqn\cyouf{
\hat I_0 \sim \cases{ \CU & $-{3\pi \over 4} < \arg w < {3\pi \over 4} $\cr
 \CE_0 & $-{\pi \over 4} < \arg(-w) < {\pi \over 4} $\cr}
}
Here $\CE_0 \sim \sqrt{2} w^{-2}\exp[w^2/8] $ is the 
growing exponential corresponding to the critical point with $\nu=0$ (or its LG images).
\eqn\cyouff{
\hat I_1 \sim \cases{ \CU + {2\pi i \over 4} \log w + \cdots & $-{\pi \over 4} < \arg w < \pi  $\cr
\CU - {3\pi i \over 4} \log w + \cdots & $-\pi < \arg w < -{3\pi \over 4}  $\cr
 \CE_1 & $-{3\pi \over 4} < \arg w < -{2\pi \over 4} $\cr}
}
Here $\CE_1 \sim \sqrt{2} w^{-2}\exp[-w^2/8] $ is the 
growing exponential corresponding to the critical point with $\nu=1$ (or its LG images). 
Similarly
\eqn\cyoufff{
\hat I_{-1} \sim \cases{ \CU - {2\pi i \over 4} \log w + \cdots & $-\pi  < \arg w < {\pi \over 4} $\cr
\CU +{3\pi i \over 4} \log w + \cdots & ${3\pi \over 4}  < \arg w < \pi   $\cr
 \CE_1 & $ {\pi \over 4} < \arg  w < {3\pi \over 4} $\cr}
}
\eqn\cyoufv{
\hat I_{2} \sim \cases{ \CU + i \pi  \log w + \cdots & ${\pi\over 4}   < \arg w <\pi  $\cr
\CU -i \pi  \log w + \cdots & $-\pi   < \arg w < -{\pi \over 4}    $\cr
 \CE_0 & $ -{\pi \over 4} < \arg  w < {\pi \over 4} $\cr}
}

Using these formulae we can specify bases for the 3-step filtration $\CF_- \subset \CF_0 \subset \CF_+$, where 
$\CF_-$ has at most exponential decay, and $\CF_0$ has at most logarithmic growth. In the 
sectors $\vert \arg w \vert < {\pi \over 4} $ we find that $\CF_- $ is generated by 
$2 \hat I_0 - \hat I_1 - \hat I_{-1} \sim cnst. \CE_1$, while $\CF_0$ is spanned by 
$\hat I_0 - \hat I_1$ and $\hat I_0 - \hat I_{-1}$. The filtrations in the other sectors are given by 
applying the LG symmetry. 

\subsec{Stokes matrices}

\def\CS{{\cal S}}
\lref\balser{W. Balser, W.B. Jurkat, D.A. Lutz, {\it Birkhoff Invariants 
and
Stokes' Multipliers for Meromorphic Linear Differential Equations},
Journal of Math. Analysis and Applications, {\bf 71}(1979)48-94}

\lref\dingle{R.B. Dingle, {\it Asymptotic Expansions: their derivation and
interpretation}, Academic Press, 1973}

The differential equation \gkzsimp\ can be written as a first order 
$n\times n$ matrix equation
of the form:
\eqn\matrixfrm{
\bigl({d\over dw} - A(w) \bigr) \Psi =0
}
where
\eqn\apto{
A(w) = \sum_{i=1}^{n-1} e_{i,i+1} + {(-1)^n\over n^2} w e_{n,2} + 
{(-1)^n\over n^2} w^2 e_{n,3}
}
and $e_{i,j}$ are matrix units. 
We regard this as an equation for an $n\times n$ invertible matrix $\Psi$.
The physical interpretation of $\Psi$ is that it is the matrix of 1-point 
correlators of the elements of the chiral ring $\bigl(\IC[x_1,x_3]/(x_i \p_i W)\bigr)^{\IZ_n}$:
\eqn\chiralrng{
\Psi_{ij} = \int_{\gamma_i} (x_1 x_3)^{j} e^{-W} {dx_1\over x_1} {dx_3 \over x_3} 
}
At infinity there is a formal asymptotic solution. True $n \times n$ matrix
solutions asymptotic to the fixed  formal solution can only be defined 
in angular sectors.
For sectors of sufficiently wide angle the true solution is unique. On 
overlapping
sectors two such solutions will be related by right multiplication by a 
constant matrix
known as a Stokes matrix.  For further details see, for examples, 
\dingle\balser.

The results of the previous sections allow one to determine the Stokes 
matrices for
$n=3,4$. 
We work directly with a basis for $\cal V$, rather then $\Psi$.
For $n=3$ we define a vector of formal solutions
\eqn\formalsl{
\pmatrix{ \kappa_0 b_0 \cr \kappa_1 b_1 \cr}
}
where $\kappa_0,\kappa_1$ are appropriate constants and $b_0, b_1$ are 
defined in \asympbasis.
We now introduce
six sectors: $\CS_j:= \{ w\vert {\pi j\over 3} - {\pi \over 2}
< \arg w < {\pi j\over 3} + {\pi \over 6} \},\, 1\leq j\leq6$. In each of these sectors 
there is a {\it unique}
basis of solutions asymptotic to \formalsl. These are:
\eqn\trusol{
\eqalign{
\CS_1:\qquad\qquad & \psi_1 = \pmatrix{ \hat I_0 - \hat I_1 \cr 2 \hat 
I_0 - \hat I_1 - \hat I_{-1} \cr} \cr
\CS_2: \qquad\qquad & \psi_2 = \pmatrix{ \hat I_0 - \hat I_1 \cr 2 \hat 
I_1 - \hat I_0 - \hat I_{-1} \cr} \cr
\CS_3: \qquad\qquad & \psi_3  = \pmatrix{ \hat I_1 - \hat I_{-1} \cr 2 
\hat I_1 - \hat I_0 - \hat I_{-1} \cr} \cr
 \cdots\qquad\qquad\,\,&\qquad\qquad\qquad\cdots \cr}
}
The other sectors are obtained by LG symmetry. Then we have Stokes 
matrices:
\eqn\sotks{
\eqalign{
\CS_1 \cap \CS_2:\qquad\qquad  & \psi_1 = \pmatrix{1&0 \cr 3 & 1\cr} 
\psi_2 \cr
\CS_2 \cap \CS_3: \qquad\qquad & \psi_2 = \pmatrix{1&-1\cr 0& 1\cr} 
\psi_3 \cr
\cdots\qquad\qquad\quad\,\,&\qquad\qquad\,\cdots \cr}
}
and the remaining sectors are obtained by applying LG symmetry.
Similarly, one can compute the Stokes matrices for $n=4$.

\subsec{Summary}

The generalized periods are solutions of the GKZ equation \gkzsimp. A basis of solutions can be written in 
the integral representation \lgi\ \inthat\ for appropriate linear combinations of  regulated ``straightline contours'' \gami. 
 These linear combinations should 
be thought of as elements of the homology group $H_n(\IC^2, B)$ where $B$ is the region at 
infinity where $ \re W \to +\infty$. By combining the ``Landau-Ginzburg symmetry'' 
\lgsymm\ with saddle point techniques ( equations \lgiiip\lgivp\spcontr) and the integral 
representation \newrep\ we are able to find the angular-sector-dependent asymptotic behavior of the 
generalized periods, which exhibit Stokes' phenomenon.
We gave a detailed analysis   for the cases of $n=3,4$ in \calee--\cyoufv\ and showed how 
to compute the Stokes' matrices in \sotks.

\newsec{Propeller branes vs. fractional branes: analysis for small $n$}
In this section we analyze $\IC^2/\IZ_{n(p=1)}$ orbifolds for 
$n=3,4$.
Let us first summarize our findings.
We look at the overlap of the Ramond ground state with the branes
in the theory deformed by the addition of a tachyon vertex operator.
This quantity is the direct analog of the central charge in
the spacetime supersymmetric theories and we call it the ``generalized
central charge'' or ``generalized period.''
Earlier in the paper we explained that this quantity satisfies
the GKZ equation.
A convenient basis of non-constant solutions is provided by 
combinations of $\hat I_s$ which satisfy  \newslkc.

As $x_1,x_3\ra \infty$, the A-brane surfaces defined in Section 3 
asymptote to quarterplanes \gami\ used
in the definition of $I_{\alpha,\beta}$ in \regin.
More precisely, we find that 
\eqn\cgccp{  \langle c_\nu|1\rangle = {\cal C}_1 +
                  {\cal C}_2 (\hat I_{s+1}+\hat I_{s-1}-2\hat I_s)       }
with the relation between $\nu$ and $s$ to be determined later.
In eq. \cgccp\ ${\cal C}_1$ and ${\cal C}_2$ 
are $w$-independent constants.
We do not know the behavior of the contour $\Gamma_\nu$ in the 
compact region. Because of the pole $1/x_1 x_3$ in the 
measure we must allow for a   constant term  $C_1$ 
which cannot be determined by our techniques\foot{
Recall that the intersection matrix is also insensitive to
the addition/subtraction of D0 branes, which corresponds to
adding/subtracting an integer from \cgccp.}.
In the following, we will adopt the notation 
\eqn\cgcc{  \langle c_\nu|1\rangle \sim \hat I_{s+1}+\hat I_{s-1}-2\hat I_s     }
which is meant to be equivalent to \cgccp.
The evidence for \cgcc\ comes from numerically solving the differential
equations that define the profile of the A-brane.
[We also observe that \cgcc\ is consistent with the saddle-point evaluation
of the integral \gcc.]

The theory at the orbifold point contains a set of fractional branes 
whose boundary states are described in appendix A.
In the following we will provide evidence that the coulomb branch branes
are equivalent to the fractional branes, up to a permutation.
In fact, we propose
\eqn\fbgcc{  \langle e_a|1\rangle ={\cal C}_2( \hat I_{a+1}+\hat I_{a-1}-2 \hat I_a)+{1\over n}  }
(From the exact boundary state description we happen to know the
value of the constant term for this overlap; it is equal to $1/n$.)
The evidence for \fbgcc\ comes from the map between the higgs branch brane
and the fractional branes, which is determined by the intersection matrix.
In fact, for $n=3$ the map also implies that the coulomb branch brane is
equal to the fractional brane up to permutations.
Here is a consistency check for \fbgcc.
Consider turning on the tachyon vertex operator $w$.
The worldsheet action of the orbifold theory is modified by
\eqn\modac{  \delta S\sim w\int dz d\bar z \int d^2\theta X_{1\over n}    .  } 
The disk one-point function of the unit operator to   first nontrivial 
order in $w$ is
\eqn\onept{  \langle e_a|1\rangle\sim  w\, \exp\left(-{2\pi i a \over n}\right)+\OO(w^2)  }
where we used the boundary state expression for the fractional brane
to determine the one-point function, and we have dropped $n$-dependent
normalization constants.
Substituting the small $w$ expansion \itoeff\ into \fbgcc\ gives precisely
\onept.

\subsec{$n=3$}
We first consider the case of $n=3$, $p=1$.
In fact, the orbifold group is $Z_{N=2n}=Z_6$ due to the action
on fermions; only the type 0 (not the type II) theory can be defined.
There are three choices for the intersection matrix for fractional branes, given by \inonsym\
and by matrices obtained from \inonsym\ by permutation of columns.
\item{(1)} The complete, unreduced intersection matrix, obtained by a 
permutation of columns from \inonsym\ is 
\eqn\ithreeanr{ \I=  \pmatrix{~-2& ~1& ~1\cr ~1& ~-2& ~1\cr ~1& ~1& -2}   }
The reduced matrix, with the D0 brane factored
out, corresponds to the continued fraction
determined by $n/(n-p)\ra[2,2]$.
\eqn\ithreea{ \It=  \pmatrix{~-2& ~1\cr ~1& ~-2 }   }
The corresponding \HJ\ space is an ALE space. It is obtained from the orbifold
by turning on the generators of the (c,c) ring.
This is essentially equivalent to the spacetime supersymmetric case.
There is no coulomb branch and the map between the fractional
branes and the higgs branch branes is known via the McKay correspondence.
\item{(2)} The unreduced matrix is obtained by a permutation of 
columns in \ithreeanr.
The reduced intersection form is determined by the continued fraction
determined by $n/p\ra [3]$.
\eqn\ithreeb{ \It=  \pmatrix{~1& ~-2\cr ~1& ~1 }   }
According to \abc, the higgs and the coulomb branch branes are given by
\eqn\hbbthree{  h=e_s-e_{s+1},\qquad c=e_{s+1}    }
where the index $s$ is not determined at this stage.
In fact, as we will see below, all permutations will be realized,
depending on the argument of $w$.
\item{(3)} Similar to the previous case, 
the continued fraction is $n/p\ra [3]$ and
\eqn\ithreec{ \It=  \pmatrix{~1& ~1\cr ~-2& ~1 }   }
and therefore
\eqn\hbbthreeb{  h=e_s+2e_{s+1},\qquad c=e_{s+1}    }

\noindent
To proceed further, it will be convenient to have a map of the $w$-plane
divided into angular sectors.
This is shown in Fig. 3.
\midinsert\bigskip{\vbox{{\epsfxsize=3in
        \nobreak
    \centerline{\epsfbox{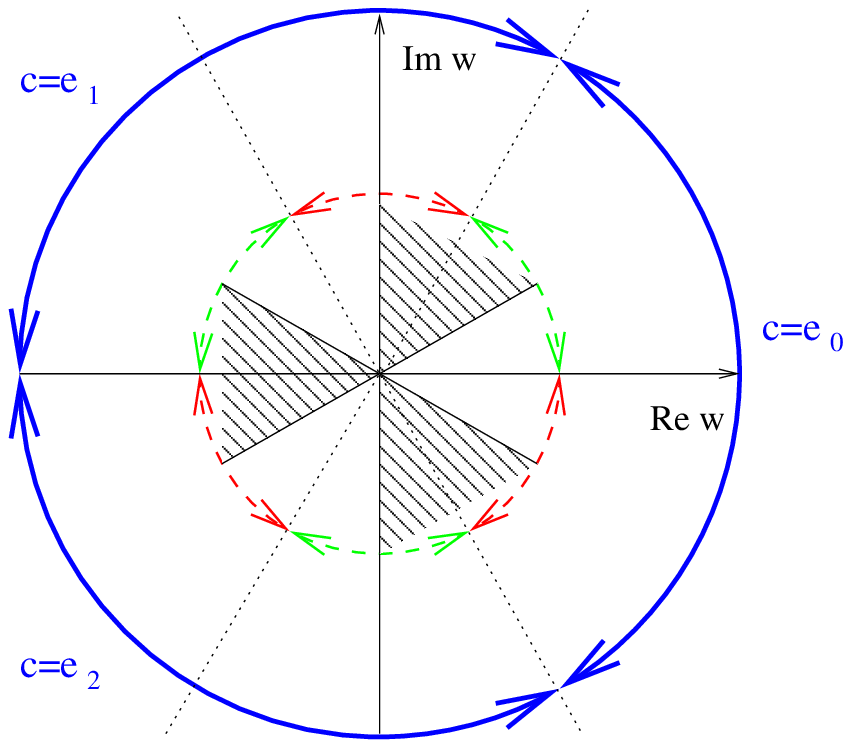}}
        \nobreak\bigskip
    {\raggedright\it \vbox{
{\bf Fig 3.}
{\it The angular structure in the $w$ plane. $w$ in the shaded sectors
gives rise to a critical point with a negative value of $Re\, W$.
Big blue arrows mark sectors which differ by permutations of 
fractional branes. Dashed red and green arrows mark sectors which differ
by the type of higgs branch brane}}}}}}
\bigskip\endinsert
According to \lgiiip, modulo the LG symmetry, there is a single critical point of the
superpotential $W$.
Its coordinates, and the value of $W$ are
\eqn\cpthree{  x_1=x_3=-{w\over 3},\qquad W_*=\left({w\over 3}\right)^3  }
Hence there are three angular sectors in the $w$-plane 
where $W_*$ is negative and the integral \gcc\ can pick up 
an exponentially growing contribution from the critical point.
This happens when $(x_1,x_3)$ given by \cpthree\ fall into the
regions (see \expnsg\ above)
\eqn\dashthree{
\eqalign{ {5\pi\over 6}\leq \arg(w)&\leq {7\pi\over 6}, \qquad  s=0   \cr
          {3\pi\over 2}\leq \arg(w)&\leq {11\pi\over 6}, \qquad s=1   \cr
          {\pi\over 6}\leq \arg(w)&\leq {\pi\over 2}, \qquad s=-1     \cr
          }
}
These are the shaded sectors in Fig. 3.
The value of $s$ in \dashthree\ specifies which $\hat I_s$ receives
an exponentially growing contribution, in accord with the rule \insectr.
In these sectors, the nonconstant solutions of the GKZ equation are spanned by the 
two functions with the following leading asymptotics:
\eqn\gkzthree{ 
\eqalign{  {5\pi\over 6}\leq \arg(w)&\leq {7\pi\over 6}:\quad \hat I_{-1}-\hat I_1\sim {2\pi i\over3}\log w,\quad
                                                        \hat I_0-I_{-1}\sim \exp(-W_*) \cr
       {3\pi\over 2}\leq \arg(w)&\leq {11\pi\over 6}:\quad \hat I_0-\hat I_{-1}\sim {2\pi i\over3}\log w,\quad
                                                         \hat I_1-I_{0}\sim \exp(-W_*) \cr
       {\pi\over 6}\leq \arg(w)&\leq {\pi\over 2}:\quad \hat I_1-\hat I_{0}\sim {2\pi i\over3}\log w,\quad
                                                         \hat I_{-1}-I_{1}\sim \exp(-W_*) \cr }
}
where $W_*$ is given by \cpthree\ and is a growing exponential.
The precise form of the exponential growth is given in \calee, but
we use $e^{-W_*}$ as a shorthand for $\CE$.
Note that the three sectors are different by a simple permutation of
indices of $\hat I$'s.
We will see below that this is a general phenomenon which is rooted in
the permutation symmetry of the fractional branes.

Consider now three sectors where $Re W_*$ is positive, and the contribution
from the nonzero critical point is a decaying exponential.
These are the unshaded sectors in Fig. 3.
The leading contribution now comes from the critical point at $(x_1,x_3)=(0,0)$
and is logarithmic.
\eqn\gkzthreeb{ 
\eqalign{  -{\pi\over 6}\leq \arg(w)&\leq {\pi\over 6}:\quad \hat I_{1}-\hat I_0\sim {2\pi i\over3}\log w,
                   \quad  \langle c|1\rangle\sim \hat I_1+\hat I_{-1}-2\hat I_0\sim \exp(-W_*) \cr
       {\pi\over 2}\leq \arg(w)&\leq {5\pi\over 6}:\quad \hat I_{-1}-\hat I_1\sim {2\pi i\over3}\log w,
                  \quad  \langle c|1\rangle\sim \hat I_{-1}+\hat I_{0}-2\hat I_1\sim \exp(-W_*) \cr
       {7\pi\over 6}\leq \arg(w)&\leq {3\pi\over 2}:\quad \hat I_{0}-\hat I_{-1}\sim {2\pi i\over3}\log w,
                  \quad  \langle c|1\rangle\sim  \hat I_{0}+\hat I_{1}-2\hat I_{-1}\sim \exp(-W_*) \cr }
} 
In eq. \gkzthreeb\ we identified the exponentially decaying solution 
with the generalized central charge of the coulomb branch brane.
Indeed, we expect the latter to receive an exponentially decaying contribution
from the critical point of $W$.
Now both \hbbthree\ and \hbbthreeb\ state that the coulomb branch brane
is the same as the fractional brane, up to a permutation.
But we can fix the freedom with the help of \onept.
This leads to \fbgcc.

The picture that we infer from \gkzthreeb\ is therefore the following.
In the (undashed) sector  $-{\pi\over 6}\leq \arg(w)\leq {\pi\over 6}$, $c=e_0$.
Rotating $\arg(w)$ by $2\pi/3$ enforces the permutation of fractional branes.
E.g. in the sector ${\pi\over 2}\leq \arg(w)\leq {5\pi\over 6}$,  $c=e_1$ etc.
What we do not yet know is how the transition between $c=e_0$ and $c=e_1$
(and more generally, between  $c=e_s$ and $c=e_{s+1}$) happens.
A natural scenario would be the following.
The $w$ plane is divided into 3 angular regions, marked by the blue arrows in Fig. 3.
In the region $-\pi/3\leq\arg(w)\leq\pi/3$, $c=e_0$, and $w\ra \exp(2\pi i /3)w$
enforces the permutation of the fractional branes $e_s\ra e_{s+1}$.
Unfortunately we cannot see this directly from the asymptotics of $\hat I$'s.
The reason is that both $e_s$ and $e_{s+1}$ have the same exponentially
growing asymptotics in the region where the transition happens.
However there is a way to verify the picture proposed above.
As explained before, the shape of the coulomb branch brane can be inferred from
the solution of the soliton equations.
The corresponding solutions emanate from the critical point of $W$ and
run to infinity along the quarterplanes $\gamma_{\alpha\beta}$ used
to define the $\hat I_{\alpha\beta}$'s
(there are $n=3$ inequivalent choices of them).
Solving the equations near the critical point and far away, in the asymptotic region,
is easy.
The more difficult question is matching the solutions in the two regions, i.e. 
understanding which direction in the $w$ plane is chosen by a certain soliton trajectory,
and how this choice depends on $\arg (w)$.
In appendix B we analyze this question numerically.
Our analysis confirms the picture described above.
That is, there are three angular sectors in the $w$ plane, and
the asymptotics of the A-brane surface emanating from the nonzero critical point
(i.e. coulomb branch brane) jump as $w$ crosses the lines of $\arg(w)=\pi/3,\pi,-\pi/3$.
Moreover, the asymptotics are consistent with \gkzthreeb.

So far we have determined the behavior of the generalized central charge for the coulomb
branch brane as a function of $w$.
The theory contains a single higgs branch brane, which wraps 
the exceptional $\IP^1$  of $\OO(-3)\ra \IP^1$.
Unfortunately we cannot analyze the soliton equations for this brane,
since the LG model is unreliable near $(x_1,x_3)=(0,0)$ which would
serve as a critical point.
Nevertheless, we can use \hbbthree\ and \hbbthreeb\ together with \fbgcc\
to compute the generalized central charge.
Let us specialize the discussion to the angular sector $-\pi/6\leq\arg(w)\leq\pi/6$;
the other two sectors differ by permutation.
In this sector $c=e_0$, therefore the two possibilities for the higgs branch brane
consistent with the intersection matrix  are
$h^{(1)}=e_2-e_0$ and $h^{(2)}=e_2+2 e_0$.
(Again, one must keep in mind that the expression for the the higgs and
coulomb branch branes are obtained modulo the addition/subtraction
of D0 branes)
Using \fbgcc\ we obtain 
\eqn\hbgccthree{  \langle h^{(1)}|1\rangle\sim3(\hat I_0-\hat I_{-1}),\quad                  
                  \langle h^{(2)}|1\rangle\sim3(\hat I_1-\hat I_{0})       }
We can identify the logarithmic solution with the higgs branch
brane for the shaded sectors in the $w$ plane.
This is where it is defined uniquely (in the unshaded sectors, 
one can add an exponentially decaying solution without changing 
the leading asymptotics).
Comparing with \gkzthree\ we observe that for $-\pi/3\leq\arg(w)\leq-\pi/6$,
$h^{(1)}=e_2-e_0$, while for $\pi/6\leq\arg(w)\leq\pi/3$, $h^{(2)}=e_2+2e_0$.
The natural line of transition between the two happens at $\arg(w)=0$.
(This would be a direct analog of the transition between $c=e_s$ and $c=e_{s+1}$).
We cannot rigorously prove this point and leave
further justification to the future\foot{ 
We can solve the soliton equations {\it with the flat metric}
similarly to what is described in appendix B.
The solutions emanating from the $(x_1,x_3)=(0,0)$ critical point
suggest that the corresponding A-brane surface indeed asymptotes to 
the quarter planes defining $3(\hat I_0-\hat I_{-1})$ and $3(\hat I_1-\hat I_0)$
and the transition between the two happens at $\arg(w)=0$.
This is another strong piece of evidence in favor of the picture proposed.}.

Let us summarize.
There are $n=3$ ``big'' angular sectors.
In the sector $-\pi/3\leq\arg(w)\leq\pi/3$, the coulomb branch 
brane $c=e_0$.
Multiplication by $\exp(2\pi i/3)$ causes cyclic permutations
of the fractional branes $e_0\ra e_1$ etc.
In addition, each big sector is divided into two smaller ones.
For example, in the angular sector $-\pi/3\leq\arg(w)\leq\pi/3$
there are two possible higgs branch branes:
\eqn\hbfbthreeb{
\eqalign{    h^{(1)}&=e_2-e_0+\ell D0, \quad -{\pi\over3}\leq\arg(w)\leq0   \cr
              h^{(2)}&=e_0-e_1+\ell' D0, \quad 0\leq\arg(w)\leq{\pi\over3}   \cr}
}
where $\ell$ and $\ell'$ are nonnegative integers.
It is natural to assume that $\ell'=\ell$.
The two branes in \hbfbthreeb\ then differ by a permutation.
Note that both higgs branch branes in \hbfbthreeb\ have the same
asymptotic behavior for the generalized central charge as a function
of $w$:
\eqn\gcchbbt{   \langle  h^{(1,2)}|1\rangle\sim 2\pi i\log w   }
This is reminiscent of the mirror map for the spacetime non-supersymmetric case.
When $\ell=0$ or $\ell'=0$, the branes in \hbfbthreeb\ become massless at the orbifold point, which
generally signals the breakdown of the string perturbation theory.
It is possible however, that the states which appear in the Hilbert space
of the orbifold theory have positive $\ell,\ell'$.

\subsec{n=4}
Our next example is $n=4$, $p=1$.
In this case the orbifold group is  $Z_{N=n}=Z_4$ 
and both type 0 and type II theory can be defined. 
There are $n=4$ choices for the reduced intersection matrix $\It$.
\item{(1)} The intersection form corresponds to the continued fraction
determined by $n/(n-p)\ra[2,2,2]$.
\eqn\ifoura{ \It=  \pmatrix{~-2& ~1& ~0\cr ~1& ~-2& ~1\cr ~0& ~1& ~-2 }   }
This describes spacetime-supersymmetric ALE space.
In all other examples the intersection form is determined by the continued fraction
determined by $n/p\ra [4]$.
\item{(2)} 
\eqn\ifourb{ \It=  \pmatrix{~1& ~-2& ~-1\cr ~0& ~1& ~-2\cr ~1& ~0& ~1 }   }
The higgs branch brane is given by
\eqn\hbbfour{  h^{(1)}=e_s+2 e_{s+1}+3 e_{s+2}    }
\item{(3)} 
\eqn\ifourc{ \It=  \pmatrix{~0& ~1& ~-2\cr ~1& ~0& ~1\cr ~-2& ~1& ~0 }   }
and therefore
\eqn\hbbfourb{  h^{(2)}=e_s+2 e_{s+1}- e_{s+2}    }
In fact, $\It$ in \ifourc\ is the intersection matrix for the 
type II string, and \hbbfourb\ is the unique higgs branch brane
in this case.
\item{(4)} 
\eqn\ifourd{ \It=  \pmatrix{~1& ~0& ~1\cr ~-2& ~1& ~0\cr ~1& ~-2& ~1 }   }
This is the transpose of \ifourb.
The higgs branch brane is 
\eqn\hbbfourd{  h^{(3)}=e_s-2 e_{s+1}- e_{s+2}    }

\noindent
We also claim that in $(2)-(4)$
\eqn\cbbfour{ c_1=e_{s+1},\, c_2=e_{s+2},\quad {\rm or}\quad 
          c_1=e_{s+2},\, c_2=e_{s+1}   }
The reasoning goes as follows.
Below we will see that the generalized central charge for
the coulomb branch branes is given by \cgcc.
This formula satisfies (in a certain angular sector)
\eqn\ccnu{  \langle c_k|1\rangle\sim w\, \exp\left({2\pi i (k-2)\over n}\right)   +\OO(w^2),
\qquad k=1,2} 
The intersection form implies that $c_k$ and $e_k$ are possibly
related as
\eqn\cerel{   \pmatrix{~e_s\cr ~e_{s+1}}=\pmatrix{~a& ~b\cr ~c& ~d} \pmatrix{~c_1\cr ~c_2},\qquad
     ad-bc=1                      }
The consistency with \onept\ implies that in this angular sector
$e_0=c_2,e_1=c_1$ and, more generally, \cbbfour.
(The alternative solution is $e_1=c_1,e_2=-c_2$ implies ``negative mass'' for the 
$c_1$ brane and therefore should be discarded.
It is also not consistent with the formula for the higgs branch brane.)

What happens as the phase of $w$ is varied?
The relevant angular sectors are depicted in Fig. 4.
There are two critical points (modulo LG symmetry)
\eqn\critfour{ \nu=0:\;\; (x_1,x_3)=( {\sqrt{-w}\over 2},{\sqrt{-w}\over 2});\qquad
  \nu=1:\;\; (x_1,x_3)=( {\sqrt{-w}e^{\pi i \over4}\over 2},{\sqrt{-w}e^{3\pi i\over4}\over 2})} 
The choice of sign in the square root does not matter, the difference
amounts to the LG symmetry.
We choose a fundamental domain to be $0\leq\arg(-w)<2\pi$ which maps into 
$0\leq\arg(\sqrt{-w})<\pi$.
For a given value of $w$, $Re W_*$ has different signs for the two critical
points.
One critical point contributes a growing exponential, while the other a decaying
one.
The $w$ plane is divided into four big sectors, which can labeled by
$\nu=0,1$ and $s=0,1,2,3$, depending on which critical point contributes
a growing exponential to which $\hat I_s$.
\midinsert\bigskip{\vbox{{\epsfxsize=3in
        \nobreak
    \centerline{\epsfbox{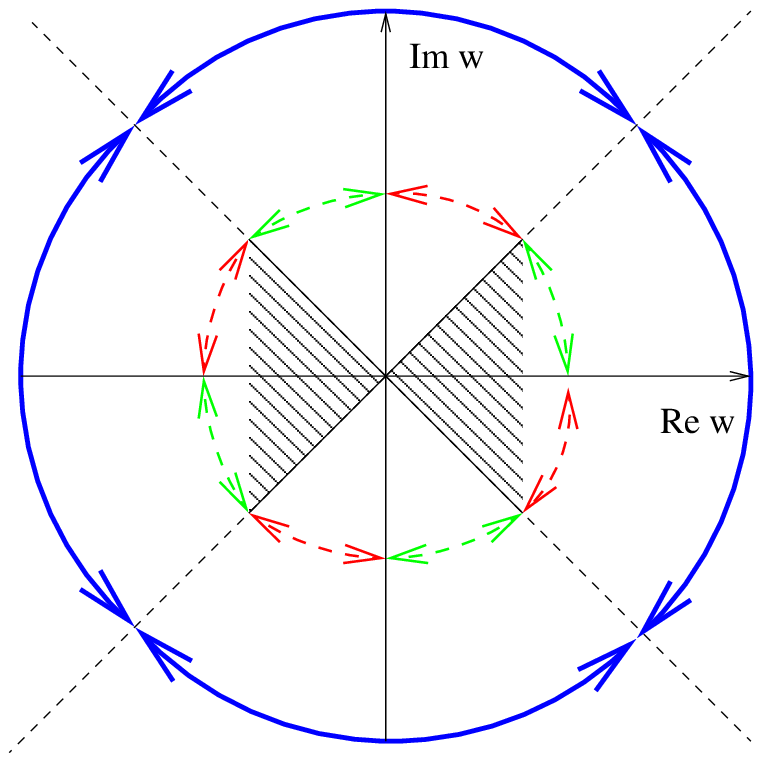}}
        \nobreak\bigskip
    {\raggedright\it \vbox{
{\bf Fig 4.}
{\it The angular structure in the $w$ plane for $n=4$. When $w$ is in the shaded sectors,
the $\nu=0$ ($\nu=1$)critical point has negative (positive) value of $Re\, W$.
This is reversed in the unshaded sectors.
Big blue arrows mark sectors which differ by permutations of 
fractional branes. Dashed red and green arrows mark sectors which differ
by the type of higgs branch brane}}}}}}
\bigskip\endinsert
\eqn\nus{  
\eqalign{   -{\pi\over4}&< \arg(w)<{\pi\over4}:\quad \nu=0,\;s=2\cr
            {\pi\over4}&< \arg(w)<{3\pi\over4}:\quad \nu=1,\;s=3\cr
            {3\pi\over4}&< \arg(w)<{5\pi\over4}:\quad \nu=0,\;s=0\cr
            {5\pi\over4}&< \arg(w)<{7\pi\over4}:\quad \nu=1,\;s=1\cr}
}
These big sectors differ from each other by a permutation 
of fractional branes, in exact analogy to the $n=3$ case.
The exponentially decaying solution can be identified unambiguously;
the logarithmically growing solution is defined up to an addition of exponentially
decaying piece:
\eqn\logexpf{ 
\eqalign{   -{\pi\over4}&< \arg(w)<{\pi\over4}:\quad \hat I_1-\hat I_3\sim{\pi i\over2}\log w,\quad
                                            \hat I_1+\hat I_3-2\hat I_0\sim \exp(-W_1)  \cr
            {\pi\over4}&< \arg(w)<{3\pi\over4}:\quad \hat I_2-\hat I_0\sim{\pi i\over2}\log w,\quad
                                            \hat I_2+\hat I_0-2\hat I_1\sim \exp(-W_0)  \cr
            {3\pi\over4}&< \arg(w)<{5\pi\over4}:\quad \hat I_3-\hat I_1\sim{\pi i\over2}\log w,\quad
                                            \hat I_3+\hat I_1-2\hat I_2\sim \exp(-W_1)  \cr
            {5\pi\over4}&< \arg(w)<{7\pi\over4}:\quad \hat I_0-\hat I_2\sim{\pi i\over2}\log w,\quad
                                            \hat I_2+\hat I_0-2\hat I_1\sim \exp(-W_0)  \cr}
}
Let us restrict to the big sector $ -{\pi\over4}< \arg(w)<{\pi\over4}$.
The exponentially decaying solution is given by the first equation
in \logexpf.
According to \fbgcc, the respective coulomb branch brane $c_2$ must 
be equal to $e_0$.
Hence, there are two possibilities, consistent with \cbbfour.
Either $c_1=e_1$, $c_2=e_0$ and $s=3$ in \cbbfour\ (remember that $s$ is
defined mod $n$.)
Then, $\langle c_2|1\rangle\sim \hat I_1+\hat I_3-2\hat I_0$,
$\langle c_1|1\rangle\sim \hat I_2+\hat I_0-2\hat I_1$.
In this case
\eqn\casea{
\eqalign{  \langle h^{(1)}|1\rangle&\sim 4(\hat I_2-\hat I_1)\sim \exp(-W_0)   \cr
           \langle h^{(2)}|1\rangle&\sim 4(\hat I_1-\hat I_0)\sim 2\pi i \log(w)   \cr
           \langle h^{(3)}|1\rangle&\sim 4(\hat I_0-\hat I_3)\sim 2\pi i \log(w)   \cr}
}
The generalized central charge for $h^{(1)}$ has a growing exponential
behavior instead of the expected logarithmic behavior.
Alternatively, it might happen that 
$c_1=e_3$, $c_2=e_0$, $s=2$ in \cbbfour, and
hence $\langle c_1|1\rangle\sim \hat I_0+\hat I_2-2\hat I_3$,
$\langle c_2|1\rangle\sim \hat I_1+\hat I_3-2\hat I_0$ and
\eqn\caseb{
\eqalign{  \langle h^{(1)}|1\rangle&\sim 4(\hat I_1-\hat I_0)\sim 2\pi i \log(w)  \cr
           \langle h^{(2)}|1\rangle&\sim 4(\hat I_0-\hat I_3)\sim 2\pi i \log(w)   \cr
           \langle h^{(3)}|1\rangle&\sim 4(\hat I_3-\hat I_2)\sim \exp(-W_0)   \cr}
}
Now it is $h^{(3)}$ whose behavior is not consistent with the expectations.

In the Appendix B we analyze the behavior of the A-brane surfaces.
The results suggest that the sector $-\pi/4\leq\arg(w)\leq\pi/4$ is
subdivided into the two subsectors,
similar to the $n=3$ case.
Whenever $0\leq\arg(w)\leq\pi/4$, we have $c_2=e_0$, $c_1=e_1$,
and $\langle h|1\rangle\sim 4(\hat I_1-\hat I_0)$.
That is, in this subsector 
\eqn\hbfbfoura{ h=h^{(2)}=e_3+2e_0-e_1, \qquad 0\leq\arg(w)\leq {\pi\over4}  }
Whenever $-\pi/4\leq\arg(w)\leq 0$, we have $c_1=e_3$, $c_2=e_0$,
and $\langle h|1\rangle\sim 4(\hat I_0-\hat I_3)$.
In this subsector we again have
\eqn\hbfbfourb{ h=h^{(2)}=e_2+2e_3-e_0, \qquad -{\pi\over4}\leq\arg(w)\leq 0 }
Hence, among the three possible expressions for the
higgs branch brane consistent with the intersection form, 
only one is realized.
It is the one which appears in the type II case.
[As before, there is a freedom to add D0 branes to \hbfbfoura, \hbfbfourb]

In Appendix B we also analyze the shape of the brane emanating from
the $\nu=1$ critical point.
In the analysis above it has been assumed that in the
angular sector $-\pi/4\leq\arg(w)\leq\pi/4$, the corresponding
central charge is given by $\langle c|1\rangle\sim \hat I_1+\hat I_3-2 \hat I_0$,
since the value of $W$ at the critical point is positive and
we expect the generalized central charge to decay exponentially.
This is confirmed by solving the soliton equations in Appendix B.

\newsec{Discussion}
In this paper we studied A-branes in the LG model which
describes the resolution of the spacetime non-supersymmetric $\IC^2/\IZ_{n(p=1)}$
orbifold.
The model has coulomb branch branes, supported at the minima
of the superpotential $W$ away from the origin, and higgs branch 
branes, associated with the critical point at the origin.
The generalized central charge for these branes, defined in Section 3, 
satisfies the GKZ equation.
This fact, together with a knowledge of the critical points of $W$,
and the numerical analysis of the A-brane shape, allows
us to compute the central charges for both coulomb and higgs branch branes.
The results have the expected asymptotic behavior (determined by the
value of $W$ at the corresponding critical point).
They are also consistent with the open string Witten index.
That is, in the examples that we have studied, we can 
identify coulomb branch branes and fractional branes using
the intersection matrix and first order conformal
perturbation theory at the orbifold point.
The higgs branch brane is given by a linear combination
of fractional branes, so its generalized central charge
can be computed accordingly.
This procedure gives the same result as integrating
\gcc\ over the higgs branch brane propeller surface\foot{
Numerical data suggests that the propeller surface
depends only weakly  on the LG metric: its asymptotics,
which are important for the value of the integral, seem
to be metric-independent.}.
This result has logarithmic asymptotics as $|w|\ra\infty$,
as expected from the form of the integrand at the
origin.
Similar logarithmic behavior arises also in the
spacetime supersymmetric case.
  
The complex $w$ plane, is divided into $n$ angular 
sectors, related by the permutation of the fractional branes.
Hence, it is the phase of the tachyon VEV that is responsible 
for a particular ordering of fractional branes being realized.
Moreover, each sector is further divided into subsectors, where
different expressions for the higgs branch brane are valid.

In writing formulae for the higgs branch brane one must bear in mind that
our techniques are not powerful enough to distinguish between $h$ and
$h  + \ell D0$ where $\ell \in \IZ$ and $D0 = e_0 + \cdots + e_{n-1}$.
Recall that $D0$ is in the annihilator of the intersection form.
In principle this ambiguity could be fixed by determining the
constant $C_*$ in $\langle h \vert 1 \rangle = C_* + \sum c_s \hat I_s$
using the propeller surface to determine \gcc. Unfortunately we have
not been able to extract this constant. In the case $n=3$ we found
the combination $h=e_2 - e_0$ in \hbfbthreeb\ when $\ell=0$. Taken at face value
this would be a massless brane at the orbifold point, signaling a
breakdown in string perturbation theory. For this reason we find
positive values of $\ell$ in \hbfbthreeb\ more likely. Clearly, further work is needed here.

Among the coulomb branch branes, approximately one half 
have exponentially growing periods while the other half have exponentially
decaying  periods.
The block-diagonalization of the
intersection matrix implies that the coulomb branch branes are
decoupled from their higgs branch counterparts.
Surprisingly, the two coulomb branch branes in $\IC^2/\IZ_{4(1)}$
are not orthogonal to each other, even though the coulomb
vacua are far separated in the IR!
It would be interesting to analyze in detail the patterns
at higher values of $n$ where many coulomb branch branes
associated with critical points with both positive and 
negative values of $Re\, W$ are present.
This might be useful for the physical interpretation of \gcc\ in the non-conformal 
case.

This work raises a number of technical issues which must be 
solved in order to make further progress. 
We used numerical analysis to determine the shape of
the propeller branes.
The wings depend on the angular sectors in the $w$ plane;
the asymptotics change discontinuously as $w$ crosses the
borders of these sectors.
It would be nice to have some analytic technology to
understand these phenomena better.
In the $n=4$ case, some wings developed which did not seem
to contribute to \gcc.
This will probably be a persistent issue for
higher $n$, and  understanding better the shape of the propeller
brane and \gcc\ is important for making progress. Similarly, 
we have made an important assumption that the asymptotics of the 
propeller branes is independent of the choice of metric 
$g_{i\bar j}$ used in \soleq. We have checked this numerically 
for the metrics of interest here (see appendix B) but some 
rigorous results concerning this would be most welcome. 
Finally, in constructing a  map between the fractional
branes and the coulomb branch branes, it is would be very useful
to know the intersection form for the latter.
 
Understanding the structure of branes in more general
spacetime non-supersymmetric orbifolds is another interesting
direction.
At present, the higgs branch brane can be expressed in
terms of fractional branes for the $\IC^2/\IZ_{n(p)}$ orbifolds
which have a resolution in type II theory \mm.
Generalizing this result to type 0 theory for arbitrary
$n$ and $p$ will probably involve understanding phase
diagrams of multiple tachyons, and the space
of possible higgs branch branes.
It would also be worthwhile understanding spacetime nonsupersymmetric
$\IC^3/\IZ_n$ orbifolds, where some new features appear already in
the closed string sector \refs{\MNP\SarkarRY-\MorrisonJA}.

The Stokes' phenomenon observed in this paper is very likely related to 
that associated to general
semisimple Frobenius manifolds in \dubrovinI\dubrovinII. It is possible 
that some of the techniques
used in \guzzetti\ueda\ can be applied to elucidate the behavior of 
generalized periods and their
Stokes matrices for general $\IC^2/\IZ_{n(1)}$ orbifolds, or even 
$\IC^2/\IZ_{n(p)}$ orbifolds.
We hope our considerations will be useful in understanding homological 
mirror symmetry for non-Fano manifolds. In the Fano case Stokes' 
matrices are related to the dimensions of Ext groups of exceptional collections 
in the derived  category of the mirror \guzzetti\ueda. 
(For recent progress in homological mirror symmetry in the Fano case see 
\arouxI\arouxII.) In the examples of the  $\IC^2/\IZ_{n(p)}$ orbifolds 
we are instead trying to formulate a quantum version of the McKay 
correspondence, as explained in \mm\MP. One point which is currently 
missing is an  analogous interpretation of the Stokes' matrices
in the non-Fano case.

Stokes' phenomenon has recently played an important role in brane physics in the 
context of minimal string theory \MaldacenaSN. It is interesting to contrast 
that application with the present one. In both cases a ``brane partition function'' 
satisfies a differential equation with an irregular singular point, and the 
angular-sector-dependence of exponential growth and decay has important
physical consequences. In both cases one can  use branes to probe the nature 
of spacetime, and Stokes' phenomenon has important implications for  the resulting 
spacetime picture. On the other hand, in \MaldacenaSN\ one works with the 
{\it non-perturbative} brane amplitude (the Baker-Akhiezer function of the matrix 
model, now also known as the ``FZZT partition function'' ) but the present paper 
only makes use of the {\it perturbative} disk one-point function.  Nevertheless, 
there is a common mathematical thread in both examples, since  both examples 
are governed by a family of Landau-Ginzburg theories, and hence by a similar 
 underlying structure of a Frobenius manifold.

In conclusion there is an amazingly rich structure in the D-branes of the
$\IC^2/\IZ_{n(1)}$ orbifolds. 
Surely it is even more intricate in the  $\IC^2/\IZ_{n(p)}$ and
$\IC^n/\Gamma$ orbifolds.
Elucidating this structure appears to be a challenging project.

\bigskip\bigskip\noindent{\bf Acknowledgements:}
We would like to thank E. Diaconescu, R. Karp and R. Plesser 
for discussions.
We also thank B. Florea for initial participation in the project
and for many useful discussions.
This work was supported in part by  DOE grant DE-FG02-96ER40949.

\appendix{A}{Boundary states for fractional branes}
In this appendix we describe in more detail the
construction of the boundary states which correspond to fractional 
branes at the $\IC^2/\IZ_N$ orbifold.
There are two cases: (1) $p$ odd, $N=n$; (2) $p$ even, $N=2n$.
A useful reference is \bcr.
The first step is constructing B-type Ishibashi boundary states
\eqn\ishia{
\eqalign{
     (\alpha_{m-\nu}-{\tilde\alpha}_{-m+\nu})|s;\eta\rangle\rangle_{NSNS,RR}&=0,
                \qquad m=1,2,\ldots,\quad \nu={s\over n}   \cr
     ({\bar\alpha}_{m+\nu}-{\tilde{\bar\alpha}}_{-m-\nu})|s;\eta\rangle\rangle_{NSNS,RR}&=0,
                \qquad m=0,1,\ldots,\quad    \cr}
}
and 
\eqn\ishib{
\eqalign{ (\psi_{r-\nu}+i\eta {\tilde\psi}_{-r+\nu})|s;\eta\rangle\rangle_{NSNS,RR}&=0,
                \qquad r={1\over 2}+\IZ(\IZ)\quad {\rm for\; NS(R) },\; r-\nu\geq 0  \cr
          ({\bar\psi}_{r+\nu}+i\eta {\tilde{\bar\psi}}_{-r-\nu})|s;\eta\rangle\rangle_{NSNS,RR}&=0,
                \qquad  r+\nu\geq 0,\quad  \eta=\pm1 \cr}
}
One can verify that conditions \ishia\ and \ishib\ give rise to
the B-boundary states of the $\NN=(2,2)$ superconformal theory:
\eqn\abstates{
\eqalign{ 
                (G_r^++i\eta {\tilde G}_{-r}^+)|s;\eta\rangle\rangle_{NSNS,RR}&=0,\quad
                (G_r^-+i\eta {\tilde G}_{-r}^-)|s;\eta\rangle\rangle_{NSNS,RR}=0    \cr
                (J_n+{\tilde J}_{-n})|s;\eta\rangle\rangle_{NSNS,RR}&=0              \cr}
}
In the open string sector the basic ingredients are the characters
with the insertion of the group element $g^s$ where $g$ is the generator of $\IZ_n$
\eqn\chars{ \chi_{NS,R}^{s}(q_o)=\tr_{NS,R}\, g^s q_o^{L_0-{c\over 12}},\qquad
          \chi_{NS,R}^{(-)^F;s}(q_o)=\tr_{NS,R}\, (-)^F g^s q_o^{L_0-{c\over 12}}   }
Here $s$ runs from $1$ to $N$ and $q_o=\exp(2\pi i \tau_o)$.
When $p$ is even and $N=2n$, the number of independent characters
can be reduced to $n$ by the following identity
\eqn\idchars{  \chi_{NS}^{s+n}(q_o)= \chi_{NS}^{s}(q_o),\qquad 
             \chi_{NS}^{(-)^F;s+n}(q_o)= \chi_{NS}^{(-)^F;s}(q_o)    }
and
\eqn\idcharsb{  \chi_{R}^{s+n}(q_o)= -\chi_{R}^{s}(q_o),\qquad 
             \chi_{R}^{(-)^F;s+n}(q_o)= -\chi_{R}^{(-)^F;s}(q_o)    }
These identities follow from \orbaction.
In the $NS$ sector $g^n=1$, while in the $R$ sector $g^n=(-)^{p+1}=-1$.
The characters \chars\ have the following modular transformation properties 
under $\tau_o\ra \tau_c=-1/\tau_o$:
\eqn\mtchar{
\eqalign{ \chi_{NS}^{s}(q_o)&=\sigma(s)\; {}_{NSNS}\langle\langle s;\pm|
           q_c^{{1\over 2}(L_0+{\tilde L}_0-{c\over 12})} |s\pm\rangle\rangle_{NSNS} \cr
        \chi_{NS}^{(-)^F;s}(q_o)&=-\sigma(s)\; {}_{RR}\langle\langle s;\pm|
           q_c^{{1\over 2}(L_0+{\tilde L}_0-{c\over 12})} |s\pm\rangle\rangle_{RR} \cr
        \chi_{R}^{s}(q_o)&=\sigma(s)\; {}_{NSNS}\langle\langle s;\pm|
           q_c^{{1\over 2}(L_0+{\tilde L}_0-{c\over 12})} |s\mp\rangle\rangle_{NSNS} \cr
        \chi_{R}^{(-)^F;s}(q_o)&=-\sigma(s)\; {}_{RR}\langle\langle s;\pm|
           q_c^{{1\over 2}(L_0+{\tilde L}_0-{c\over 12})} |s\mp\rangle\rangle_{RR} \cr}
}
where 
\eqn\sigmadef{ \sigma(s)=4\sin\left({\pi s\over n}\right)\sin\left({\pi s p\over n}\right)  }
With the exception of the factor $\sigma(s)$, the modular transformation properties
\mtchar\ are the same as those of the untwisted characters.
The boundary states are constructed by requiring open-closed string duality.
Consider $N=n$ case first.
\eqn\ocdual{
\eqalign{  \tr_{ab;NS}\; q_o^{L_0-{c\over 12}}&=
            {1\over n}\sum_{s=0}^{n-1} \omega_n^{(a-b)s} \sigma(s) {}_{NSNS}\langle\langle s;\pm|
           q_c^{{1\over 2}(L_0+{\tilde L}_0-{c\over 12})} |s\pm\rangle\rangle_{NSNS}         \cr
  &={}_{NSNS}\langle a;\pm|q_c^{{1\over 2}(L_0+{\tilde L}_0-{c\over 12})} |b;\pm\rangle_{NSNS}  \cr}  
}
where we introduced the Cardy state $|a;\eta\rangle_{NSNS}$ and $\omega_n=\exp(2\pi i/n)$.
Eq. \ocdual\ and its counterpart with $\tr_{ab;R}\; q_o^{L_0-{c\over 12}}$ in the RHS implies
\eqn\cardya{|a;\eta\rangle_{NSNS}=\sum_{s=0}^{n-1}  \omega_n^{a s} \sqrt{\sigma(s)}\; 
           |s;\eta\rangle\rangle_{NSNS}                     }
Similarly,
\eqn\cardyb{|a;\eta\rangle_{RR}=\sum_{s=0}^{n-1}  \omega_n^{a s} \sqrt{\sigma(s)}\; 
           |s;\eta\rangle\rangle_{RR}                     }
Type 0 theory admits two types of branes distinguished by the value of $\eta$:
\eqn\tzerob{ |a;+\rangle={1\over\sqrt{2}}(|a;+\rangle_{NSNS}+|a;+\rangle_{RR}),\quad
             |a;-\rangle={1\over\sqrt{2}}(-|a;-\rangle_{NSNS}+|a;-\rangle_{RR})      }
In type II theory, only one combination is invariant under the GSO projection
\eqn\typetwob{ |a;II\rangle={1\over 2}
           (|a;+\rangle_{NSNS}-|a;-\rangle_{NSNS}+|a;+\rangle_{RR}+|a;-\rangle_{RR})       }
The situation with $p$ even $n$ odd, where type II can not be defined and
the orbifold group is $\IZ_N=\IZ_{2n}$ is a little bit more tricky.
Now $a$ in the open string sector is forced to run from $0$ to $2n-1$ and the analog of 
\ocdual\ is
\eqn\ocduala{   \tr_{ab;NS}\; q_o^{L_0-{c\over 12}}=
            {1\over 2n}\sum_{s=0}^{n-1} (1+(-)^{a-b})
           \omega_{2n}^{(a-b)s} \sigma(s) {}_{NSNS}\langle\langle s;\pm|
           q_c^{{1\over 2}(L_0+{\tilde L}_0-{c\over 12})} |s\pm\rangle\rangle_{NSNS}         }
where we converted the sum over $s=0,\ldots,2n-1$ to the sum over   $s=0,\ldots,n-1$
using \idchars\ and $\omega_{2n}=\exp(2\pi i/2 n)$.
The closed string sector in type 0 is invariant under $s\ra s+n$, that
is why there are $n$ Ishibashi states in \ocduala.
The sum in \ocduala\ is zero unless $a-b$ is even.
This prompts us to introduce two types of branes: the ones with $a=2a';\;\eta=+1$
and the ones with $a=2a'+1;\;\eta=-1$.
(Now, in addition to $\IZ_n$ quantum symmetry which permutes the branes, 
there is a $\IZ_2$ symmetry which amounts to multiplying $\eta$ by $-1$.) 
The boundary states are
\eqn\cardyc{|a';+\rangle_{NSNS,RR}=\sum_{s=0}^{n-1}  \omega_n^{a' s} \sqrt{\sigma(s)}\; 
           |s;+\rangle\rangle_{NSNS,RR}                     }
and
\eqn\cardyd{|a';-\rangle_{NSNS,RR}=\sum_{s=0}^{n-1}  \omega_n^{(a'+{1\over2}) s} \sqrt{\sigma(s)}\; 
           |s;-\rangle\rangle_{NSNS,RR}     }
These boundary states are consistent with \ocduala\ and its counterpart with 
$(-)^F$ inserted in the LHS.
They are also consistent with the corresponding open string expressions
in the $R$ sector:
\eqn\ocdualb{   \tr_{ab;R}\; q_o^{L_0-{c\over 12}}=
            {1\over 2n}\sum_{s=0}^{n-1} (1+(-)^{a-b-1})
           \omega_{2n}^{(a-b)s} \sigma(s) {}_{NSNS}\langle\langle s;+|
           q_c^{{1\over 2}(L_0+{\tilde L}_0-{c\over 12})} |s-\rangle\rangle_{NSNS}         }
and
\eqn\ocdualc{   \tr_{ab;R}\; (-)^F q_o^{L_0-{c\over 12}}=
            {1\over 2n}\sum_{s=0}^{n-1} (1+(-)^{a-b-1})
           \omega_{2n}^{(a-b)s} \sigma(s) {}_{RR}\langle\langle s;+|
           q_c^{{1\over 2}(L_0+{\tilde L}_0-{c\over 12})} |s-\rangle\rangle_{RR}         }

\appendix{B}{Shape of A-branes}
In this appendix we analyze the shape of A-branes in the LG theory.
As explained in \HIV\ these branes are Lagrangian surfaces
whose image in the $W$ plane is a semi-infinite real line emanating
from the critical point $\phi^i=\phi^i_*$ and going in the positive real direction.
More practically, one is instructed to solve the soliton equation \soleq.
 The shape of the A-brane associated with a given critical point is
the set of all trajectories satisfying \soleq.
Near the critical point, the set of solutions can be parametrized
by a small sphere, as in \sphere.

Let us specialize to our LG model.
The superpotential is
\eqn\spotl{ W=x_1^n+w x_1 x_3+x_3^n  }
To write the soliton equations we must make a choice of metric. The measure factor 
in \gcc\ (which originates from a path integral) suggests that one should use the 
metric $ds^2 = \vert{dx_1 \over x_1} \vert^2 + \vert{dx_3 \over x_3} \vert^2$ for 
large $x_i$. This metric is inconvenient for displaying the results of the numerical 
analysis because the soliton equations develop singularities at finite values of $\sigma$. 
We have instead used the metric $ds^2 = \vert{x_1 dx_1  } \vert^2 + \vert{x_3 dx_3  } \vert^2$ 
because this is the simplest metric for which there is no singularity at finite $\sigma$. 
A tedious numerical check shows that the asymptotics of the propeller branes described below 
is in fact independent of the choice of the metric. It would be very useful to establish 
this rigorously. Some further comments on metric dependence can be found below. 

The soliton equations are
\eqn\soleqlg{   {dx_1\over d\sigma}={1\over|x_1|^2}(n \bar x_1^{n-1}+\bar w \bar x_3),\qquad
   {dx_3\over d\sigma}={1\over|x_3|^2}(n \bar x_3^{n-1}+\bar w \bar x_1)}
The non-zero critical points are given by \lgiiip\ and the value of $W_*$
by \lgivp.
Substituting $x_1=x_1^{(\nu)}+\delta x_1, x_3=x_3^{(\nu)}+\delta x_3$ we have
\eqn\spexp{  W=W_{\nu}+{w\over 2}\left( e^{-{2\pi i\nu\over n}}(1-n)\delta x_1^2
                  +2\delta x_1 \delta x_3  + e^{2\pi i\nu\over n}(1-n)\delta x_3^2\right)   }
\subsec{n=3}
For $n=3$ there is a single critical point with $\nu=0$, $x_1=x_3=-w/3$ and $W_0=w^3/27$.
The change of coordinates which diagonalizes \spexp\ is
\eqn\ccoord{ u={\delta x_1+\delta x_3\over\sqrt{2}},\qquad   v={\delta x_1-\delta x_3\over\sqrt{2}}  }
which recasts \spexp\ as
\eqn\spexpa{  W=W_{\nu}+|w| \exp(i\varphi_w+i\pi)\left({1\over 2}u^2+{3\over2}v^2\right)   }
where we introduced $\varphi_w=\arg(w)$.
The wavefront is a circle and can be parameterized by a single angle $\theta$
\eqn\wf{  
\eqalign{   x_1&=-{w\over3}+{\epsilon\over\sqrt{|w|}} e^{-i{\varphi_w+\pi\over 2}}\left(\cos\theta+
                 {1\over\sqrt{3}}\sin\theta\right)   \cr
             x_3&=-{w\over3}+{\epsilon\over\sqrt{|w|}} e^{-i{\varphi_w+\pi\over 2}}\left(\cos\theta-
                 {1\over\sqrt{3}}\sin\theta\right)   \cr}
}
When $x_1,x_3$ are large, the $w x_1 x_3$ term in the superpotential \spotl\
is negligible.
Therefore in this regime the 
equations for $x_1$ and $x_3$ decouple and the A-brane degenerates to the product of lines
$x_1^n\in \IR_+$ and $x_3^n\in \IR_+$, which are used to define $\hat I$'s.
Each of these surfaces is essentially a quarter plane.
There are $n^2$ choices of defining lines, but
the LG symmetry brings it down to $n$ inequivalent surfaces.
Generally, a circle parametrized by $\theta$ splits into 
several components; at large worldsheet time $\sigma$, each component 
traces some quarter plane described above.
Therefore the shape of the A-brane surfaces resembles that of a propeller,
and we call these surfaces ``propeller branes''.
The integral of \gcc\ over each wing of the propeller defines a function $\hat I_s$. 

By solving the soliton equations \soleqlg\ with the initial conditions
\wf\ we can determine which combination of  $\hat I$'s corresponds to 
a given coulomb branch brane.
The relevant differential equation can be solved by Mathematica.
In Fig. 5,6  we present a plot of $\arg(x_1), \arg(x_3)$ as functions 
of $\sigma$ for several values of $\theta$.

\midinsert\bigskip{\vbox{{\epsfxsize=3in
        \nobreak
    \centerline{\epsfbox{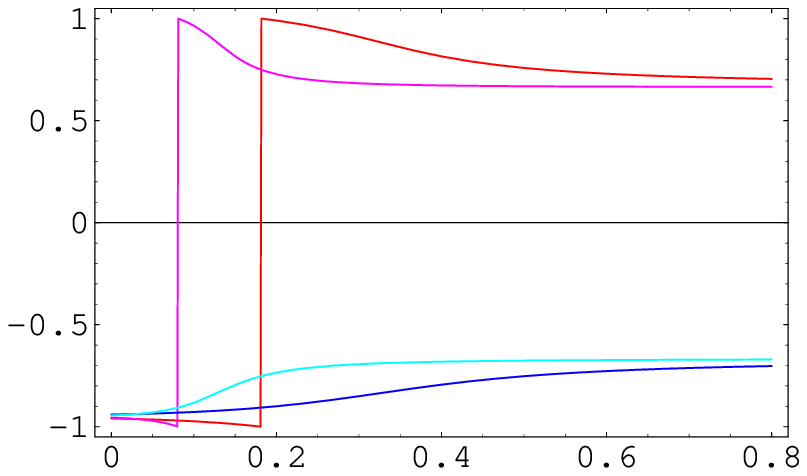}}
        \nobreak\bigskip
    {\raggedright\it \vbox{
{\bf Fig 5.}
{\it Solution of soliton equation. Horizontal axis corresponds
to $\sigma$. Vertical axis is $\arg(x_1(\sigma))/\pi$.
The lines are computed for $w=\exp(0.05 \pi i)$, $\theta=0$ (orange),
$\pi/2$ (pink), $3\pi/2$ (turquoise), $\pi$ (blue).
They asymptote to $\pm2/3$ as $\sigma\ra\infty$.}}}}}}
\bigskip\endinsert
\midinsert\bigskip{\vbox{{\epsfxsize=3in
        \nobreak
    \centerline{\epsfbox{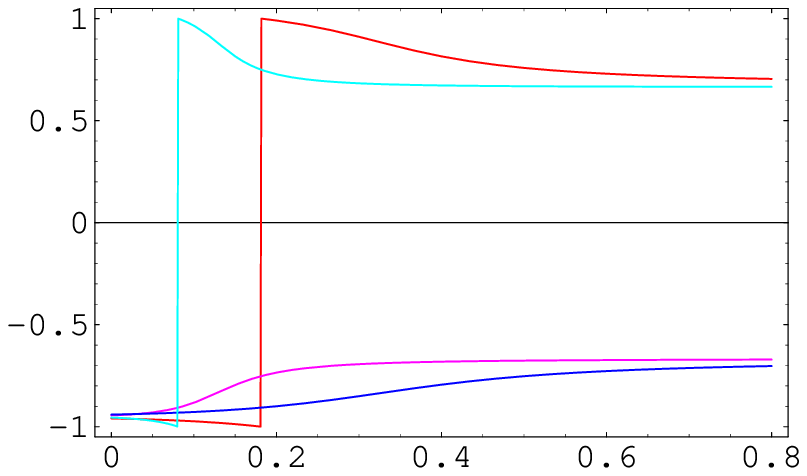}}
        \nobreak\bigskip
    {\raggedright\it \vbox{
{\bf Fig 6.}
{\it Solution of soliton equation. Horizontal axis corresponds
to $\sigma$. Vertical axis is $\arg(x_3(\sigma))/\pi$.
The lines are computed for $w=\exp(0.05 \pi i)$, $\theta=0$ (orange),
$\pi/2$ (pink), $3\pi/2$ (turquoise), $\pi$ (blue).
They asymptote to  $\pm2/3$ as $\sigma\ra\infty$.}}}}}}
\bigskip\endinsert
From Figs. 5,6 we infer that the pink [turquoise] line which corresponds to
$\theta=\pi/2$ [$\theta=3\pi/2$] asymptotes to $(\arg(x_1),\arg(x_3))=(2\pi/3,-2\pi/3)$
[ $(\arg(x_1),\arg(x_3))=(-2\pi/3,2\pi/3)$.]
For $e^{i\theta}$ in most of the upper (lower) half plane, the asymptotics are similar to those
of $\theta=\pi/2$ ($\theta=3\pi/2$).
These two regions define two different representatives of $\hat I_0$
related by the LG symmetry.
For $\theta\approx 0$ the asymptotics corresponds to $\hat I_1$,
while for $\theta\approx \pi$ we get $\hat I_{-1}$.
Hence, we observe that the behavior of solutions is consistent
with the identification 
\eqn\pbnt{ \langle c|1\rangle\sim \hat I_1+\hat I_{-1}-2 \hat I_0, \qquad  -\pi/3\leq\arg(w)\leq\pi/3 }
In fact, Figs. 5,6 do not qualitatively change in 
the whole ``big'' angular sector $-\pi/3\leq\arg(w)\leq\pi/3$.
The qualitative change happens when $w$ crosses the lines $\arg(w)=-\pi/3$
and $\arg(w)=\pi/3$.
The asymptotics of the coulomb branch brane changes in accord with 
the rule proposed above: multiplying $w$ by $\exp(2\pi i/3)$ (rotating
the ``big'' sector) corresponds to the permutation of the fractional
branes.

In the discussion above we determined that the integral \gcc\ over
the propeller surface originating at the coulomb branch critical point
gives rise to \pbnt.
This is because the propeller surface in question has four wings.
Integrating \gcc\ over these wings gives rise to $I_1$, $I_{-1}$ 
and $I_0$ (twice).
Eq. \pbnt\ is then consistent with \linde, with 
the one-point function for the fractional brane, and with
the intersection matrix.
Nevertheless it is desirable to have an independent way of
determining the orientation of the wings, i.e. the signs in
eq. \pbnt.
Consider a wing defined by $(x_1,x_3)=(e^{2\pi i s_1\over n} t_1,e^{2\pi i s_3\over n} t_3)$
for large positive $t_1$ and $t_3$.
A soliton trajectory originating at a given value of $\theta$
gives rise to a ray in $(t_1,t_3)$ plane.
In the simplest scenario, the slope of this ray $\gamma=|x_1|/|x_3|$ is a monotonic 
function of $\theta$.
The orientation is then determined by the sign of $d\gamma/d\theta$.
In Fig. 7 we plot $\gamma(\theta)$.
There are four regions where this function is monotonic; it
takes all values between zero and infinity.
These four regions correspond to  $I_1$, $I_{-1}$ 
and $I_0$ (twice).
The signs of $d\gamma/d\theta$ are consistent with the signs that appear in
\pbnt.
\midinsert\bigskip{\vbox{{\epsfxsize=3in
        \nobreak
    \centerline{\epsfbox{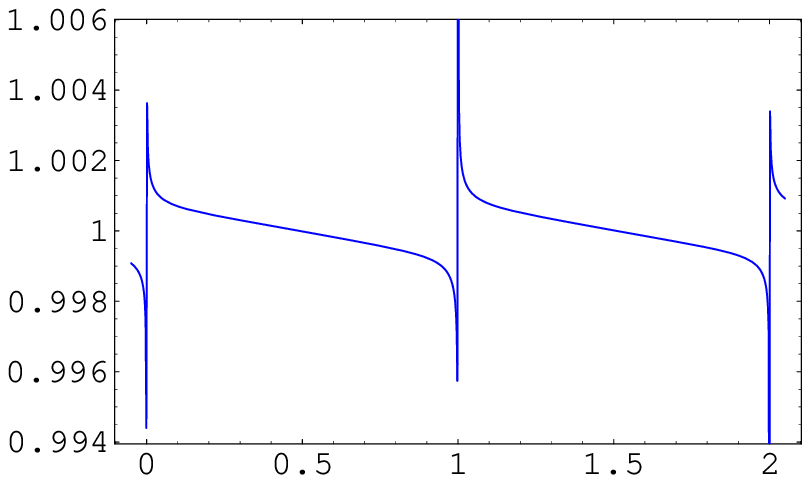}}
        \nobreak\bigskip
    {\raggedright\it \vbox{
{\bf Fig 7.}
{\it Orientation of the wings ($n=3$) is determined by $\gamma(\theta)$.
Vertical axis is $\gamma\equiv|x_1|/|x_3|$.
Horizontal axis is $\theta/\pi$.}}}}}}
\bigskip\endinsert
It is also interesting to investigate the dependence of the solutions
of soliton equations on the metric.
One natural metric to try is the Euclidean metric, leading to 
soliton equations of the form
\eqn\soleqtm{   {dx_1\over d\sigma}=(n \bar x_1^{n-1}+\bar w \bar x_3),\qquad
   {dx_3\over d\sigma}=(n \bar x_3^{n-1}+\bar w \bar x_1)}
We analyzed the solutions of these equations emanating from 
a nonzero critical point.
The technical difference with \soleqlg\ is that now $x_1$ or $x_3$
run off to infinity at finite values of $\sigma$.
Moreover, for generic values of $\theta$ only one of $x_1, x_3$
runs off to infinity, and only for isolated values of $\theta$ 
both of them do so.
All of this makes eqs. \soleqtm\ more difficult to study numerically
then eqs. \soleqlg.
The numerical data suggests that solutions of \soleqtm\ asymptote
to the surface that is homologous at the infinity to the one
defined by  \soleqlg. We have also carried out similar checks of metric 
independence for the metric 
$ds^2 = \vert{dx_1 \over x_1} \vert^2 + \vert{dx_3 \over x_3} \vert^2$ 
(this metric is the natural one to expect from the derivation of the 
LG theory via $T$-duality). 
All this suggests that the behavior of the A-brane surface depends only weakly
 on the metric, and the asymptotics might stay the same 
for a large class of metric deformations.
It is now natural to try to  analyze the behavior of the higgs branch brane as well.
The metric near the origin is not known, but if the asymptotics do not
depend on it, we might as well use eqs. \soleqtm.
The results confirm the picture described in the main text.
Restricting to the big angular sector $-\pi/3<\arg(w)<\pi/3$,
the solutions experience an abrupt change as $w$ crosses the line $\arg(w)=0$.
Above this line, the solutions asymptote to the surface defining 
$3(\hat I_1-\hat I_0)$, and below the line to $3(\hat I_0- \hat I_{-1})$.

\subsec{n=4}
Consider now $n=4$.
We will restrict our analysis to the angular sector
defined by $-\pi/4<\arg(w)<\pi/4$ (see Fig. 4).
Other big sectors are related to this one by the LG symmetry.
We studied the surface corresponding to the higgs branch brane (with the flat metric).
The result is
\eqn\hbf{ 
\eqalign{   0<\arg(w)<{\pi\over4}:\quad &\langle h|1\rangle\sim 4(\hat I_1-\hat I_0) \cr
           -{\pi\over4}<\arg(w)<0:\quad &\langle h|1\rangle\sim 4(\hat I_3-\hat I_0) \cr }
}
Consider now the contribution of the critical point with $\nu=0$,
$x_1=x_3=\sqrt{-w}/2$, $W=-2 (w/2)^2$.
(We denote the corresponding coulomb branch brane by $c_1$).
As explained before, it contributes a growing exponential, hence
it is necessary to solve the soliton equations to determine the
asymptotics.
Note that in the unshaded sectors in Fig. 4 this critical point 
contributes a decaying exponential. 
The integral \gcc\ over the coulomb branch branes associated with it,
should give rise to the following overlaps:
\eqn\oldec{
\eqalign{  {\pi\over4}<\arg(w)<{3\pi\over4}:\quad \langle c_1|1\rangle&\sim
                                   \hat I_2+\hat I_0-2 \hat I_1 \cr
{5\pi\over4}<\arg(w)<{7\pi\over4}:\quad \langle c_1|1\rangle&\sim
                                   \hat I_0+\hat I_2-2 \hat I_3 \cr}
}
Returning back to the soliton equations, the change of variable which 
diagonalizes \spexp\ is
\eqn\diagspf{ u=e^{i\varphi_w+i\pi\over2}(\delta x_1+\delta x_3),\qquad
v=e^{i\varphi_w+i\pi\over2}(\delta x_1-\delta x_3)                   }
which gives rise to
\eqn\wfourcp{W=W_0+{|w|\over 2} (u^2+2 v^2)    }
The wavefront near the critical point is 
\eqn\wffoura{    
\eqalign{   x_1&={\sqrt{-w}\over2}+{\epsilon\over\sqrt{|w|}} e^{-i{\varphi_w+\pi\over 2}}\left(\cos\theta+
                 {1\over\sqrt{2}}\sin\theta\right)   \cr
             x_3&={\sqrt{-w}\over2}+{\epsilon\over\sqrt{|w|}} e^{-i{\varphi_w+\pi\over 2}}\left(\cos\theta-
                 {1\over\sqrt{2}}\sin\theta\right)   \cr}
}
The solutions exhibit more complicated behavior than what 
we have seen so far.
As mentioned above, whenever $w$ is in the unshaded sector,
we expect the integral \gcc\ over the wings of the propeller
surface to give \oldec.
This is indeed what the numerical analysis tells us.
Suppose $\pi/4\leq\arg(w)\leq3\pi/4$.
Then integrating \gcc\ over the wings gives rise to
the first equation of \oldec.
Crossing the line $\arg(w)=\pi/4$ into the shaded sector
does not change the solution.
Figs. 8--10 contain the graphs of $\arg(x_1)$, $\arg(x_3)$
as functions of $\theta$ for large worldsheet time $\sigma$.
According to Figs. 8--10, the propeller surface emanating
from the $\nu=0$ critical point has four wings.
Integrating \gcc\ over these wings gives rise to the first
equation in \oldec.
(This is similar to the $n=3$ case)
\midinsert\bigskip{\vbox{{\epsfxsize=3in
        \nobreak
    \centerline{\epsfbox{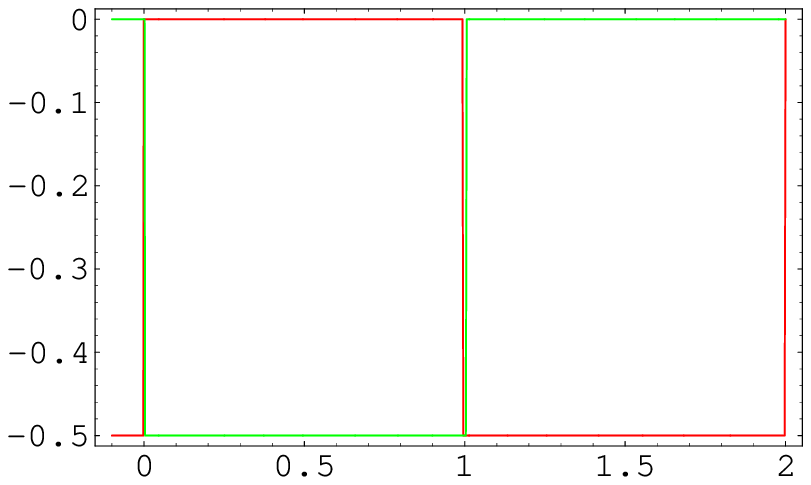}}
        \nobreak\bigskip
    {\raggedright\it \vbox{
{\bf Fig 8.}
{\it Solution of soliton equation at $\epsilon=0.005$, $\sigma=8$, $w=w^{0.9\pi i/4}$.
Horizontal axis is $\theta/\pi$. 
Vertical axis is $\arg(x_i(\sigma))/\pi$. Red line is $i=1$. Green line $i=3$.}}}}}}
\bigskip\endinsert
\midinsert\bigskip{\vbox{{\epsfxsize=3in
        \nobreak
    \centerline{\epsfbox{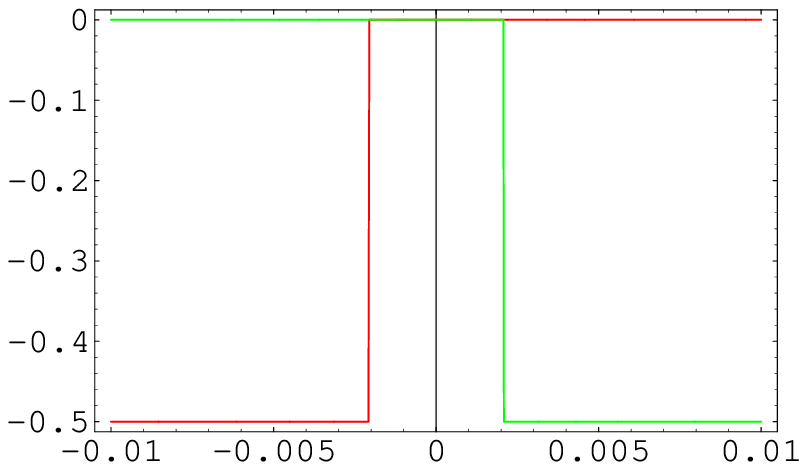}}
        \nobreak\bigskip
    {\raggedright\it \vbox{
{\bf Fig 9.}
{\it Close-up view of Fig. 8}}}}}}
\bigskip\endinsert
\midinsert\bigskip{\vbox{{\epsfxsize=3in
        \nobreak
    \centerline{\epsfbox{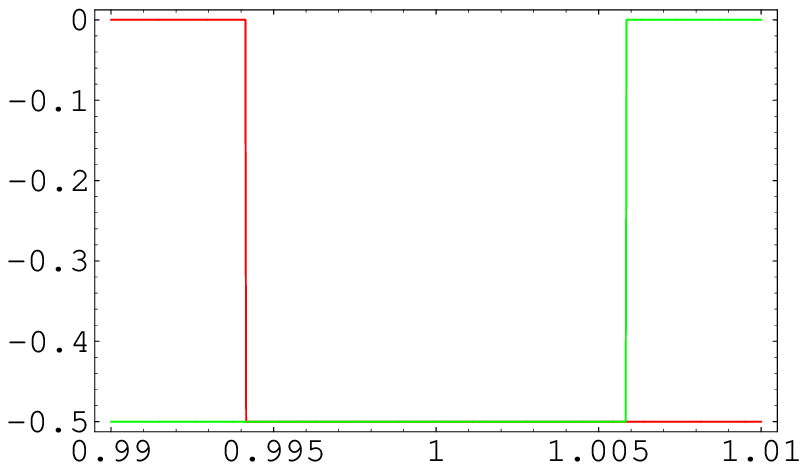}}
        \nobreak\bigskip
    {\raggedright\it \vbox{
{\bf Fig 10.}
{\it Another close-up view of Fig. 8}}}}}}
\bigskip\endinsert

To determine the orientation, we compute $\gamma(\theta)$
for large worldsheet time $\sigma$, where $\gamma\equiv |x_1|/|x_3|$,
as before.
The result is presented in Fig. 11.
It is essentially equivalent to the one in Fig. 7,
as is consistent with the signs in \oldec.
\midinsert\bigskip{\vbox{{\epsfxsize=3in
        \nobreak
    \centerline{\epsfbox{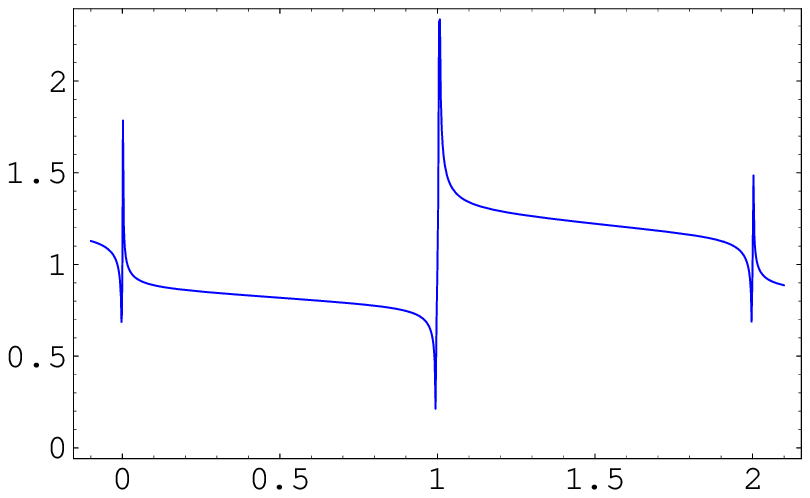}}
        \nobreak\bigskip
    {\raggedright\it \vbox{
{\bf Fig 11.}
{\it Orientation of the wings ($n=4$; $w=e^{0.9\pi i/4}$) is determined by $\gamma(\theta)$.
Vertical axis is $\gamma\equiv|x_1|/|x_3|$.
Horizontal axis is $\theta/\pi$.}}}}}}
\bigskip\endinsert
This   is not yet the end of the story, as we will see
shortly.
As $\arg(w)$ decreases past $\arg(w)\sim 0.6\pi/4$, the
form of the graphs in Fig. 8 qualitatively changes.
Fig. 12 provides an illustration for $\arg(w)=0.45\pi/4$, and
Figs. 13, 14 give the close-up views.
\midinsert\bigskip{\vbox{{\epsfxsize=3in
        \nobreak
    \centerline{\epsfbox{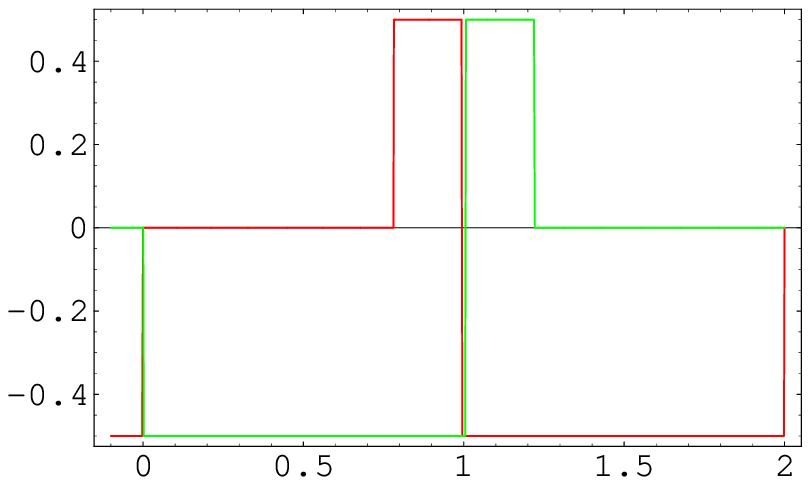}}
        \nobreak\bigskip
    {\raggedright\it \vbox{
{\bf Fig 12.}
{\it Solution of soliton equation at $\epsilon=0.005$, $\sigma=8$, $w=e^{0.45\pi i/4}$.
Horizontal axis is $\theta/\pi$. 
Vertical axis is $\arg(x_i(\sigma))/\pi$. Red line is $i=1$. Green line $i=3$.}}}}}}
\bigskip\endinsert
\midinsert\bigskip{\vbox{{\epsfxsize=3in
        \nobreak
    \centerline{\epsfbox{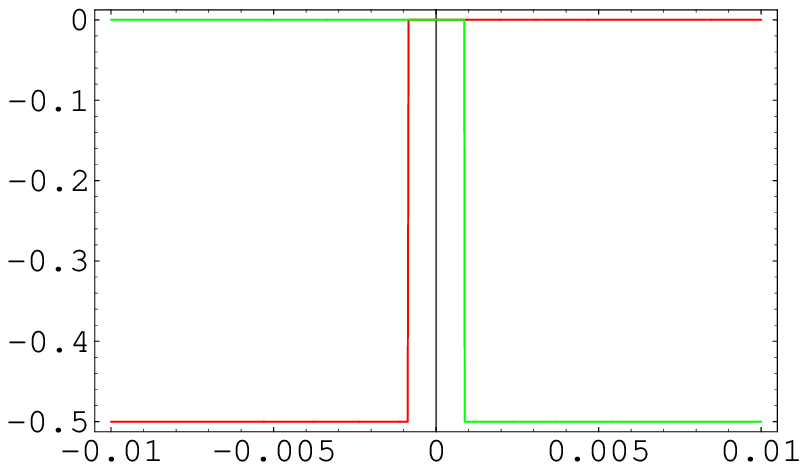}}
        \nobreak\bigskip
    {\raggedright\it \vbox{
{\bf Fig 13.}
{\it Close-up view of Fig. 12}}}}}}
\bigskip\endinsert
\midinsert\bigskip{\vbox{{\epsfxsize=3in
        \nobreak
    \centerline{\epsfbox{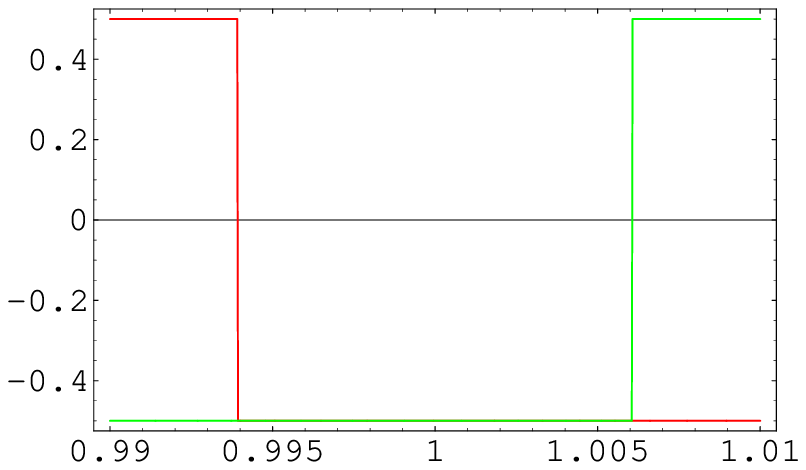}}
        \nobreak\bigskip
    {\raggedright\it \vbox{
{\bf Fig 14.}
{\it Another close-up view of Fig. 12}}}}}}
\bigskip\endinsert
Figs. 12--14 seem to suggests that the propeller surface develops two
extra wings.
The integral \gcc\ over these wings can cancel, 
so Figs. 12--14 are still consistent with the expression for the central charge 
in \oldec.
Yet, they are also consistent with $\langle1|c\rangle\sim \hat I_2+2\hat I_1-3\hat I_0$\foot{
Note that as $\arg(w)\ra 0$, the wings which give rise to $\hat I_1$ shrink
but do not disappear until $w$ crosses the $\arg(w)=0$ line.
On the contrary, the new wings, which may define extra $2 I_0$ disappear as $\arg(w)$ is
increased over $\sim0.6\pi/4$.}
To resolve this ambiguity, we look again at the $\gamma(\theta)$
graph which determines the orientation.
Corresponding graphs are presented in Figs. 15--17.
\midinsert\bigskip{\vbox{{\epsfxsize=3in
        \nobreak
    \centerline{\epsfbox{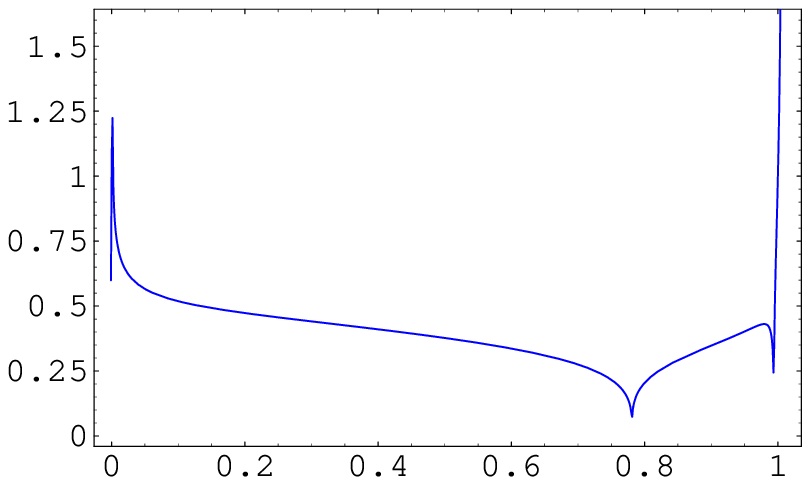}}
        \nobreak\bigskip
    {\raggedright\it \vbox{
{\bf Fig 15.}
{\it Orientation of the wings ($n=4$; $w=e^{0.45\pi i/4}$) is determined by $\gamma(\theta)$.
Vertical axis is $\gamma\equiv|x_1|/|x_3|$.
Horizontal axis is $\theta/\pi$.}}}}}}
\bigskip\endinsert
\midinsert\bigskip{\vbox{{\epsfxsize=3in
        \nobreak
    \centerline{\epsfbox{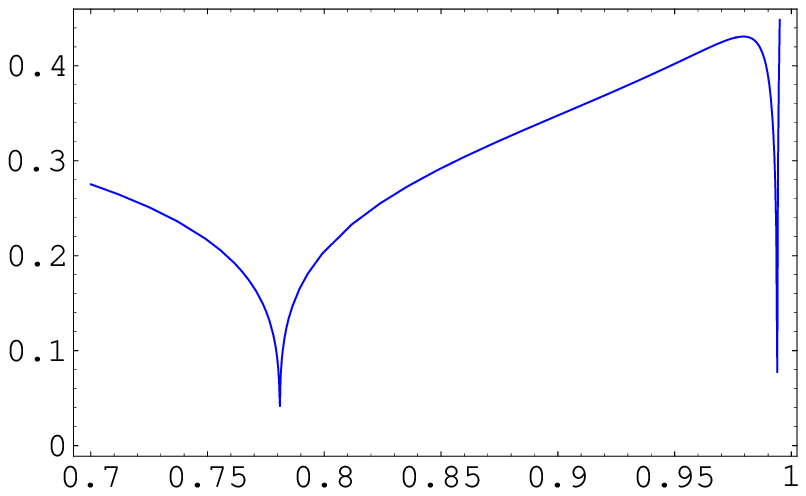}}
        \nobreak\bigskip
    {\raggedright\it \vbox{
{\bf Fig 16.}
{\it Close-up view of Fig. 15.}}}}}}
\bigskip\endinsert
\midinsert\bigskip{\vbox{{\epsfxsize=3in
        \nobreak
    \centerline{\epsfbox{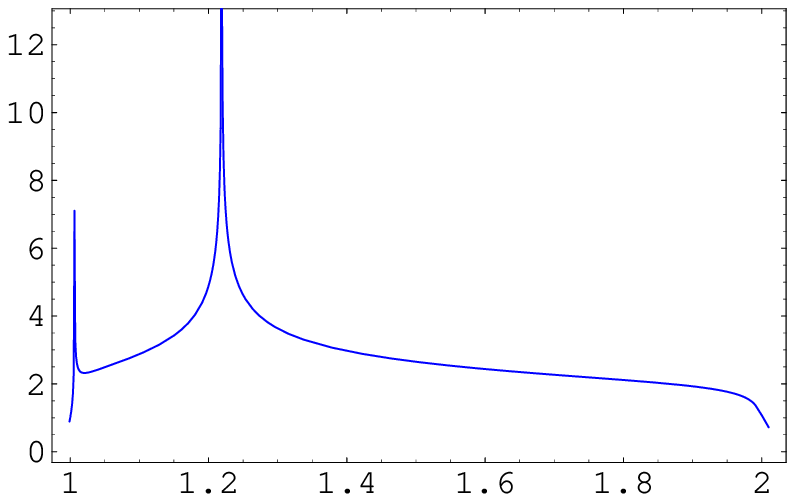}}
        \nobreak\bigskip
    {\raggedright\it \vbox{
{\bf Fig 17.}
{\it Orientation of the wings ($n=4$; $w=e^{0.45\pi i/4}$) is determined by $\gamma(\theta)$.
Vertical axis is $\gamma\equiv|x_1|/|x_3|$.
Horizontal axis is $\theta/\pi$.}}}}}}
\bigskip\endinsert
From Figs. 15--17 we immediately see that
the orientation of the four wings which appeared in 
Fig. 6 does not change as $\arg(w)$ is decreased.
Therefore, we must associate 
$\langle c_1|1\rangle \sim \hat I_2+\hat I_0-2 \hat I_1$, as before.
The integrals over the new wings must cancel each other,
instead of giving an extra $\hat I_0$.
A similar situation happens for $\arg(w)<0$.

To summarize, in the sector $-\pi/4\leq\arg(w)\leq\pi/4$ we have
[compare with \oldec]
\eqn\cbsf{
\eqalign{  0<\arg(w)<{\pi\over4}:\quad \langle c_1|1\rangle&\sim
                                   \hat I_2+\hat I_0-2 \hat I_1 \cr
-{\pi\over4}<\arg(w)<0:\quad \langle c_1|1\rangle&\sim
                                   \hat I_0+\hat I_2-2 \hat I_3 \cr}
}
This implies
\eqn\cbfbfs{
\eqalign{  0<\arg(w)<{\pi\over4}:\quad c_2=e_0; c_1=e_1 \cr
-{\pi\over4}<\arg(w)<0:\quad c_2=e_0; c_1=e_3           \cr}
}

Consider now the critical point labeled by $\nu=1$,
giving rise to the coulomb branch brane
denoted by $c_2$.
The choice of coordinates which diagonalizes \spexp\ is
now given by
\eqn\diagno{  u=e^{{i\varphi_w\over2}+{i\pi\over4}}\left[
   {1\over 2}\left( {3\over\sqrt{2}}-2\right)^{1/2}\delta x_3 
   -{i\over2}\left({3\over\sqrt{2}}+2\right)^{1/2} \delta x_1 \right] }
and
\eqn\diagnov{  v=e^{{i\varphi_w\over2}+{i\pi\over4}}\left[
   {1\over 2}\left( {3\over\sqrt{2}}+2\right)^{1/2}\delta x_3 
   -{i\over2}\left({3\over\sqrt{2}}-2\right)^{1/2} \delta x_1 \right] }
which corresponds to
\eqn\wfourcpp{W=W_1+\sqrt{2} |w| (u^2+v^2)    }
The wavefront near the critical point is
\eqn\wffourb{    
\eqalign{   x_1&={\sqrt{-w}\over2}e^{\pi i\over4}+
  {\epsilon\over\sqrt{|w|}} e^{-{i \varphi_w\over2}-{i\pi\over 4}}
    \left[  {i\over 2}\left( {3\over\sqrt{2}}+2\right)^{1/2}\cos\theta 
   -{i\over2}\left({3\over\sqrt{2}}-2\right)^{1/2} \sin\theta \right]       \cr
             x_3&={\sqrt{-w}\over2}e^{3\pi i\over4}+
  {\epsilon\over\sqrt{|w|}} e^{-{i \varphi_w\over2}-{i\pi\over 4}}
 \left[  -{1\over 2}\left( {3\over\sqrt{2}}-2\right)^{1/2}\cos\theta 
   +{1\over2}\left({3\over\sqrt{2}}+2\right)^{1/2} \sin\theta \right]  \cr}
}
In the angular sector $-\pi/4\leq\arg(w)\leq\pi/4$, the propeller
surface has four wings, and the integral \gcc\ reduces to
$\langle c_2|1\rangle\sim \hat I_1+\hat I_3-2\hat I_0$.
This is the unique exponentially decaying solution in this
sector.
This result is consistent with the fact that the critical
value of $W$ at the $\nu=1$ critical point is positive,
so we expect the corresponding central charge to decay
exponentially.

\listrefs

\end